\documentclass[12pt]{article}
\usepackage{eqsection}
\usepackage{epsf}

\def\beq{\begin{equation}}
\def\eeq{\end{equation}}
\def\bea{\begin{eqnarray}}
\def\eea{\end{eqnarray}}

\def\nn{\nonumber}

\relax
\begin{document}

\begin{titlepage}
\begin{center}
{\large\bf Parafermionic theory with the symmetry $Z_{N}$, for $N$ odd.} \\[.5in]
{\bf Vladimir S.~Dotsenko\bf${}^{(1)}$,
 \bf Jesper Lykke Jacobsen\bf${}^{(2)}$
 \bf and Raoul Santachiara\bf${}^{(1)}$}\\[.2in]
{\bf (1)} {\it LPTHE\/}\footnote{Laboratoire associ\'e No. 280 au
CNRS},
         {\it Universit{\'e} Pierre et Marie Curie, Paris VI\\
               Bo\^{\i}te 126, Tour 16, 1$^{\it er}$ {\'e}tage,
               4 place Jussieu, F-75252 Paris Cedex 05, France.}\\
{\bf (2)} {\it Laboratoire de Physique Th\'eorique et Mod\`eles
               Statistiques, \\
               Universit\'e Paris-Sud, B\^atiment 100, F-91405 Orsay,
               France.}\\[.2in]
dotsenko@lpthe.jussieu.fr, jacobsen@ipno.in2p3.fr,
santachiara@lpthe.jussieu.fr
\end{center}

\underline{Abstract.}

We construct a parafermionic conformal theory with the symmetry
$Z_N$, for $N$ odd, based on the second solution of
Fateev-Zamolodchikov for the corresponding parafermionic chiral
algebra. Primary operators are classified according to their transformation
properties under the dihedral group $D_N$, as singlet, doublet 
$1,2,\ldots,\frac{N-1}{2}$, and disorder operators. In an assumed Coulomb gas
scenario, the corresponding vertex operators are accommodated by the weight
lattice of the Lie algebra $B_{(N-1)/2}$. The unitary theories are
representations of the coset
$SO_n(N) \times SO_2(N) / SO_{n+2}(N)$, with $n=1,2,\ldots$.
Physically, they realise the series of multicritical points in
statistical theories having a $D_N$ symmetry.

\end{titlepage}

\newpage

\section{Introduction}

Parafermionic algebras are associated in a natural way with extra
discrete group symmetries in two-dimensional statistical systems.

The most used parafermionic conformal theory, which was constructed by Fateev
and Zamolodchikov in Ref.~\cite{para1}, describes the self-dual critical
points of the Z$_{N}$-invariant generalisations of the Ising model.
The values of the spins of the parafermions take the minimal possible
values admitted by the associativity constraints for the corresponding chiral
algebra. In this case, the central charge of the Virasoro algebra is fixed as
a function of $N$ only; there is only one conformal field theory for each $N$.

In the Appendix~A of Ref.~\cite{para1}, the same authors showed that there
exists another associative chiral algebra, with the next allowed values of the
spins of the parafermions. In this second solution, the central charge and the
structure constants of the algebra are given, for each $N$, as functions of a
free parameter.

The conformal field theories which correspond to this second solution are
known only for particular values of $N$.

The case $N=2$ corresponds to the superconformal theory ($\Delta=3/2$). For
$N=4$, the parafermionic algebra turns out to factorise into two independent
superconformal chiral algebras, with fields of dimensions $2$ and $3/2$. The
theory with $N=3$ has been fully developed by Fateev and Zamolodchikov in
Ref.~\cite{para2_3}. By imposing the degeneracy condition on the
representations fields, they constructed an infinite series of conformal
theories. These theories are supposed to describe the multicritical fixed
points of physical statistical systems with the $Z_3$ symmetry. Indeed, the
first theory of this series corresponds to the tricritical Potts model.

In a recent paper \cite{para2}, we presented the theory based on the second
solution of the $Z_5$ parafermionic algebra. In particular, by studying the
degenerate representations of this algebra we showed that the representation
fields are accommodated by the weight lattice of the Lie algebra $B_2$.

To obtain this result it was useful observing that the central charge of the
second solution of the $Z_{N}$ parafermionic algebra agrees with that of the
coset \cite{goddard}
\beq
 \frac{SO_{n}(N)\times SO_{2}(N)}{SO_{n+2}(N)},
 \label{SOcoset}
\eeq 
which is 
\bea
 c &=& (N-1) \left( 1-\frac{N(N-2)}{p(p+2)} \right), \label{cparamp} \\
 p &=& N-2+n. \label{paramp}
\eea
It is then natural to look for a Kac formula for the dimensions of the primary
operators based on the weight lattice of the Lie algebra $B_{\frac{N-1}{2}}$
for $N=2r+1$ (with $r \ge 2$), and of $D_{\frac{N}{2}}$ for
$N=2r$ (with $r \ge 3$).

In the present work we will present in detail the whole class of the conformal
theories based on the second solution of the $Z_{2r+1}$ parafermionic algebra.

The paper is organised as follows. In Section~2, we introduce the
parafermionic algebra presented in Ref.~\cite{para1}. We then discuss the
operator content of the theory together with the modules induced by the
primary operators. The commutation relations of the parafermionic modes in
each sector are given and some results of the degeneracy calculations are
shown. These results, together with the properties of the Coulomb gas
realisation of the theory, will be used in Section~3 to fix the values of the
boundary terms characterising each sector. Finally, in Section~4 we determine
to which sector each operator of the theory belongs. The theory we have built
is verified by considering the characteristic equations for three-point
functions. The conclusions are given in Section~5.

\section{Parafermionic algebra and representation space}

As discussed in the Introduction, we will consider the second solution of the
associative algebra presented in Appendix~A of Ref.~\cite{para1}. The operator
product expansions, defining the algebra of the parafermionic currents
$\Psi^{k}$ (with $k=1,2,\ldots,N-1$), have the form:
\bea
 \Psi^{k}(z)\Psi^{k'}(z') &=& \frac{\lambda^{k,k'}_{k+k'}}{(z-z')^{\Delta_{k}
 +\Delta_{k'}-\Delta_{k+k'}}} \nn\\
 &\times& \left \{\Psi^{k+k'}(z')+(z-z')
 \frac{\Delta_{k+k'}+\Delta_{k}-\Delta_{k'}}{2\Delta_{k+k'}}
 \partial_{z'}\Psi^{k+k'}(z')+\ldots \right \},\quad k+k'\neq 0 \nn \\
 \label{ope1}  \\
 \Psi^{k}(z)\Psi^{-k}(z') &=& \frac{1}{(z-z')^{2\Delta_{k}}}
 \left \{1+(z-z')^{2}
 \frac{2\Delta_{k}}{c}T(z')+\ldots \right \}.
 \label{ope2}
\eea

In Eqs.~(\ref{ope1})--(\ref{ope2}), the conformal dimension $\Delta_{k}$ of
the parafermionic current $\Psi^{k}$ is given by:
\beq
 \Delta_{k}=\Delta_{N-k}=\frac{2k(N-k)}{N}, \qquad
  k=\pm 1,\pm 2,\ldots,\pm \frac{N-1}{2}.
 \label{chidim2}
\eeq
Note that in the above equations and in the rest of the present manuscript,
the $Z_{N}$-charges $k$ and their sums $k+k'$ are defined modulo $N$. In
particular we write:
\beq
 \Psi^{N-k}\equiv\Psi^{-k}\equiv(\Psi^{k})^{+},\qquad \Delta_{N-k}
 \equiv\Delta_{-k}. \label{chconj}
\eeq
Within the second solution, the structure constants $\lambda_{k+k'}^{k,k'}$
and the central charge $c$ of the Virasoro algebra are given as functions of
a single free parameter $v$:
\bea
 (\lambda^{k,k'}_{k+k'})^{2} &=& \frac{\Gamma(k+k'+1)\Gamma(N-k+1)
 \Gamma(N-k'+1)}{\Gamma(k+1)\Gamma(k'+1)\Gamma(N-k-k'+1)\Gamma(N+1)}\nn\\
 &\times& \frac{\Gamma(k+k'+v)\Gamma(N+v-k)\Gamma(N+v-k')\Gamma(v)}
 {\Gamma(N+v-k-k')\Gamma(k+v)\Gamma(k'+v)\Gamma(N+v)}, \label{struct} \\
 c &=& \frac{4(N-1)(N+v-1)v}{(N+2v)(N+2v-2)}. \label{cparamv}
\eea
Changing the parametrisation $v\to n/2$, the connection between
Eq.~(\ref{cparamv}) and the coset formula (\ref{cparamp}) becomes explicit.

Among the representation fields, primaries with respect to the algebra
(\ref{ope1})--(\ref{ope2}), we expect to find singlet operators
\beq
 \Phi^{0}(z,\bar{z}) \label{sing}
\eeq
with $Z_{N}$ charge equal to $0$, and doublet $q$ operators
\beq
 \Phi^{\pm q}(z,\bar{z}), \qquad q=1,2,\ldots,\frac{N-1}{2}, \label{doubq}
\eeq
with $Z_{N}$ charge equal to $\pm q$. The singlet and doublet operators
(\ref{sing})--(\ref{doubq}) form the usual representations of the group
$Z_{N}$.

In fact, the theory we are considering is invariant under the dihedral group
$D_{N}$, which includes $Z_{N}$ as a subgroup. This can be seen from the
symmetry of Eqs.~(\ref{ope1})--(\ref{struct}) under conjugation of the $Z_{N}$
charge, $q \to N-q$, cf.~Eq.~(\ref{chconj}). The representation space should
thus include also the $N$-plet of $Z_{2}$ disorder operators, which we denote
as:
\beq
 \{R_a(z,\bar{z}),\quad a=1,\ldots,N\}.
\eeq

The presence of disorder operators in the spectrum has been shown explicitly
in the case of the first solution for general $N$ \cite{dis1}, and in the case
of the second solution for the theories with $N=3$ \cite{para2_3} and $N=5$
\cite{para2}.

The disorder fields thus play a symmetry generating role, since they complete
the cyclic group $Z_{N}$, generated by the parafermionic currents $\Psi^{\pm
k} (z)$, to the dihedral group $D_{N}$. However, the disorder fields are
non-chiral representation fields, and in particular they do not participate in
the chiral algebra \cite{para2}. The chiral algebra is therefore based on the
{\em abelian} group $Z_N$.

Until now we have specified and discussed the operator content of the theory.
The next step is to study the degenerate representations of the algebra
(\ref{ope1})--(\ref{ope2}). It is indeed well established that for a given
chiral algebra with a free parameter, the associated conformal field theory is
described by the degenerate representations of that algebra.

In order to study the degenerate representations, we first have to analyse the
structure of the modules induced by various primary fields. In
Ref.~\cite{para2} these modules were constructed for $N=5$. We shall show
below how to generalise these results to arbitrary odd $N$.

\subsection{Singlet and doublet sectors}
\subsubsection{Modules and mode commutation relations}

We start by considering the module of the identity. First of all we place the
chiral fields $\Psi^{\pm k}(z)$, with $k=1,2,\ldots,(N-1)/2$, and the
stress-energy tensor $T(z)$ in the module of the identity $I$. The levels of
these operators correspond to their conformal dimensions $\Delta_{\pm k}$. To
complete the module with the remaining levels it is sufficient to take into
account that, owing to the abelian monodromy of the fields $\Psi^{\pm k}$, the
level spacing in each sector is equal to 1.

\begin{figure}
\begin{center}
 \leavevmode
 \epsfysize=270pt{\epsffile{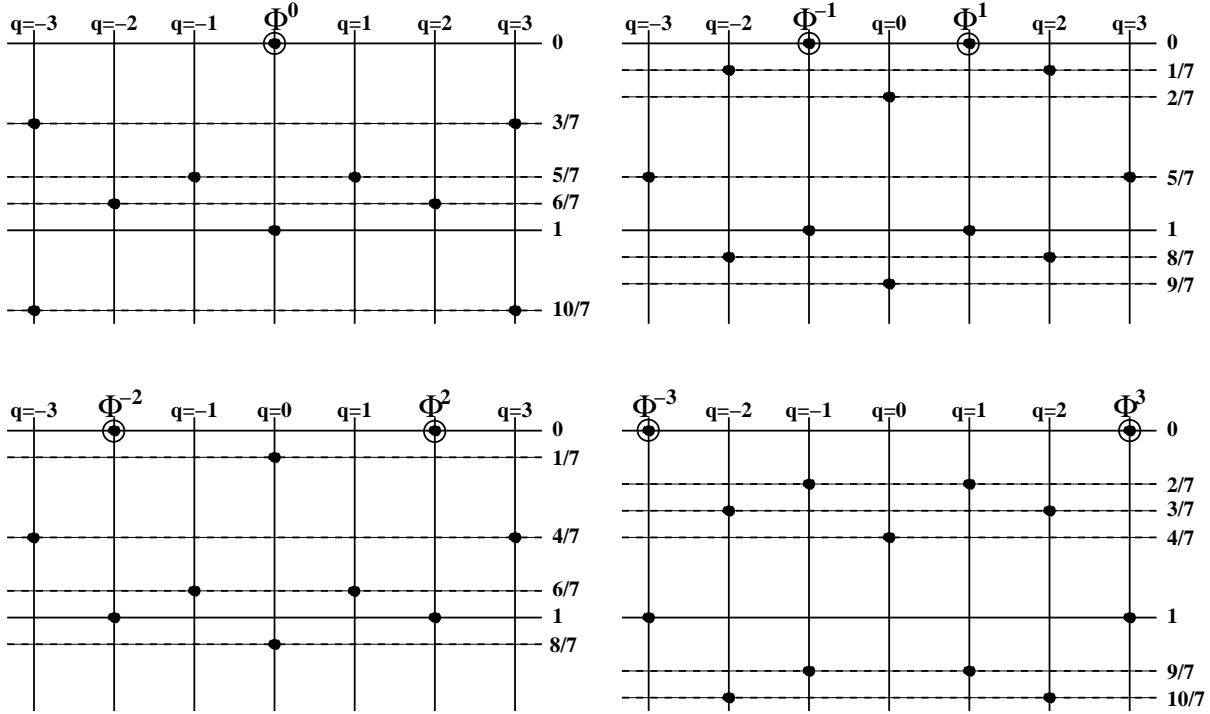}}
 \end{center}
 \protect\caption[3]{\label{fig1}Representation modules for $N=7$. We show
 the level structure of the modules of a singlet (top left), a doublet 1
 (top right), a doublet 2 (bottom left), and a doublet 3 operator
 (bottom right).}
\end{figure}

In the module of the identity there are no states below $I$ and above
$\Psi^{\pm 1}$. This is not the case for a more general singlet operator
$\Phi^{0}$; the first descendent levels of the correspondent module are, at
least partially, occupied. Let us
denote by $\delta^{0}_{k}$ the first descendent level in the $Z_N$ charge
sector $k$. {}From Eq.~(\ref{chidim2}) we readily obtain that
\beq
 \delta^{0}_{k}=-\frac{2k^2}{N} \mbox{ mod } 1. \label{levelid}
\eeq
The knowledge of the gaps $\delta^0_k$ completely fixes the level structure of
a generic singlet module. In the upper left part of Fig.~\ref{fig1} we show,
as an example, the module of a singlet operator for $N=7$.

The structure of the doublet $q$ modules is extracted from the module
of the identity by examining its corresponding submodules. The
generalisation to the doublet $q$ module of Eq.~(\ref{levelid}) is:
\beq
 \delta^{q}_{k}=2\frac{(q^2-k^2)}{N}
 \mbox{ mod } 1, \label{levelall}
\eeq
where $\delta^{q}_{k}$ is the first level in the module of the doublet $q$
corresponding to the $Z_N$ charge sector $k$. {}From now on we will make
extensive use of the above notation. In Fig.~\ref{fig1}, the three possible
doublets for $N=7$ are shown.

In accordance with the structure of the modules, the developments of the
chiral fields in each sector take, for general $N$, the form:
\bea
 \Psi^{k}(z)\Phi^{q}(0) &=&
 \sum_{n}\frac{1}{(z)^{\Delta_{k}-\delta^{q}_{k+q}+n}}
 A^{k}_{-\delta^{q}_{k+q}+n}\Phi^{q}(0), \label{development} \\
 A^{k}_{-\delta^{q}_{k+q}+n}\Phi^{q}(0) &=& 0, \qquad n>0. \label{primary}
\eea
We remind that the sums $k+q$ of the $Z_{N}$ charges are always defined modulo
$N$. Eq.~(\ref{primary}) is the usual highest weight condition which expresses
that representation fields are primary with respect to the
parafermionic algebra. As usual, from Eq.~(\ref{development}), the action
of the modes in each sector can be given in the form of a contour integral:
\beq
 A^{k}_{-\delta^{q}_{k+q}+n}\Phi^{q}(0) = \frac{1}{2\pi i}\oint_{C_{0}}
 {\rm d}z \,(z)^{\Delta_{k}-\delta^{q}_{k+q}+n-1}\Psi^{k}(z)\Phi^{q}(0).
 \label{modes}
\eeq
Note that, by Eq.~(\ref{levelall}), the level $\delta^{q}_{k+q}$ is equal to
zero for $k=-2q$. This results in the presence, for each doublet $\Phi^{\pm
q}$, of a zero mode $A^{\mp 2q}_{0}$ which acts between the two states at the
summit of the corresponding module:
\beq
 A^{\mp 2q}_{0}\Phi^{\pm q}(0)=h_{q}\Phi^{\mp q}(0). \label{zeroeigen}
\eeq
The eigenvalues $h_{q}$ defined by the above equation characterise the
representations, together with the conformal dimension of the fields
$\Phi^{\pm q}$.

The commutation relations of the mode operators can be deduced from
Eq.~(\ref{modes}) in a way similar to that used in Refs.~\cite{para2_3,para2}.
Below we give the general form of these relations.

In what follows  we define  the coefficients $D^{l}_{\alpha}$ from
the development
\beq
 (1-x)^{\alpha}=\sum^{\infty}_{l=0}D^{l}_{\alpha}x^{l}.
\eeq
By Eq.~(\ref{ope1}), the modes of the parafermions $\Psi^{k}$
and $\Psi^{k'}$, where $k,k'=\pm 1,2,\ldots,\pm \frac{N-1}{2}$ with
$k+k'\neq 0$, satisfy the relations:
\bea
 &&\sum^{\infty}_{l=0}D^{l}_{\alpha} \left(
 A^{k}_{-\delta^{k'+q}_{k+k'+q}+s(k,k',q)+n-l}
 A^{k'}_{-\delta^{q}_{k'+q}+m+l}-A^{k'}_{-\delta^{k+q}_{k'+k+q}+s(k',k,q)+m-l}
 A^{k}_{-\delta^{q}_{k+q}+n+l} \right) \Phi^{q} \nn \\
 &=&\lambda^{k,k'}_{k+k'}
 \eta^{(k,k',q)}(n,m)A^{k+k'}_{-\delta^{q}_{k+k'+q}+t(k,k',q)+n+m}\Phi^{q}
 \qquad (k+k'\neq 0) \label{qdif}
\eea
with
\bea
 \alpha &=& \Delta_{k}+\Delta_{k'}-\Delta_{k+k'} -2 \\
 \eta^{(k,k',q)}(n,m) &=&
 \Delta_{k}-\delta^{q}_{k+q}+n-1-\frac{\Delta_{k+k'}+
 \Delta_{k}-\Delta_{k'}}{2\Delta_{k+k'}}
 \left(\Delta_{k+k'}-\delta^{q}_{k+k'+q}+t(k,k',q)+n+m\right). \nn
\eea
In Eq.~(\ref{qdif}), the integers $s(k,k',q)$ and $t(k,k',q)$
shift the indices of the parafermionic modes. They are given by:
\bea
s(k,k',q)&=& \delta^{k'+q}_{k+k'+q}-\delta^{q}_{k+q}+\alpha \nn \\
t(k,k',q)&=&  \delta^{q}_{k+k'+q}-\delta^{q}_{k+q}-\delta^{q}_{k'+q}+\alpha
\eea

The remaining commutation relations between the modes of the
parafermions  $\Psi^{k}$ and $\Psi^{-k}$ (with $k=1,2,\ldots,\frac{N-1}{2}$)
realise the connection with the Virasoro algebra. In fact, it is seen from
Eq.~(\ref{ope2}) that the stress-energy operator $T(z)$ is produced
in the expansion $\Psi^{k}\Psi^{-k}$. Defining the integers $u(k,q)$
as:
\beq
 u(k,q)= \delta^{-k+q}_{q}-\delta^{q}_{k+q}+\beta
\eeq
we can write the following relations:
\bea
 &&\sum^{\infty}_{l=0}D^{l}_{\beta} \left( A^{k}_{-\delta^{-k+q}_{q}
 +u(k,q)+n-l}A^{-k}_{-\delta^{q}_{-k+q}+m+l}+
 A^{-k}_{-\delta^{k+q}_{q}+u(-k,q)+m-l}
 A^{k}_{-\delta^{q}_{k+q}+n+l} \right) \Phi^{q} \nn \\
 &=&
 \left(\kappa(n)\delta_{n+m+u'(k,q),0}+\frac{2\Delta_{k}}{c}L_{n+m+u'(k,q)}
 \right)\Phi^{q}\label{qmq}
\eea
with
\bea
 \beta &=& 2\Delta_{k}-3 \label{kmk'} \label{betadef} \\
 \kappa(n)&=&\frac{1}{2}\left(\Delta_{k}-\delta^{q}_{k+q}
 +n-1\right)\left(\Delta_{k}-\delta^{q}_{k+q} +n-2\right). \nn
\eea
On the right-hand side of Eq.~(\ref{qmq}) we have defined
$u'(k,q)=-1+u(k,q)$ for $k \neq-2q$, and $u'(k,q)=u(k,q)$ for $k=-2q$.

In Eq.~(\ref{qmq}), the $L_{n}$ are the generators of the conformal
transformations which form the Virasoro algebra:
\beq
 \left(L_{n}L_{m}-L_{m}L_{n}\right)\Phi^{q}=
 \left[(n-m)L_{n+m}+\frac{c}{12}n(n^{2}-1)\delta_{n+m,0}\right]\Phi^{q}.
 \label{virasoro}
\eeq
Finally, the list of commutation relations in the singlet and doublet sectors
are completed by the commutators between the
parafermionic modes $A^{k}$ and the $L_{n}$:
\beq
 \left(A^{k}_{-\delta^{q}_{k+q}+m}L_{n}-L_{n}A^{k}_{-\delta^{q}_{k+q}+m}\right)
 \Phi^{q}= \left[ (1-\Delta_{\Psi^{k}})n+m-\delta^{q}_{k+q} \right]
 A^{k}_{-\delta^{q}_{k+q}+m+n}\Phi^{q}. \label{parvira}
\eeq

We shall often refer to Eq.~(\ref{qdif}) using the shorthand notation
$\{\Psi^k,\Psi^{k'}\} \Phi^q$; likewise, Eq.~(\ref{qmq}) is referred
to as $\{\Psi^k,\Psi^{-k}\} \Phi^q$.
The commutation relations (\ref{qdif})-(\ref{parvira}) given above allow to
compute various matrix elements which enter the analysis of the degeneracies
in the modules, to which we now turn our attention.

In Ref.~\cite{para2}, where the case $N=5$ was considered, the
dimensions of all the fundamental operators, i.e., the fundamental
singlet, doublet 1, doublet 2 and disorder operators, were calculated
by imposing the degeneracy at the first levels of the corresponding
modules.

Below we present the analysis for the degeneracies at the first levels of the
singlet and of the doublet $1$ operators for $N=7$. In the case of the doublet
$(N-1)/2$ and of the disorder operator $R$ we have been able to treat the case
of arbitrary odd $N$. In the next Section we shall show that the results of
this analysis, together with the properties of the Coulomb gas, will allow us
to construct the theory completely.

\subsubsection{Identity operator}

In the following we shall call a degenerate operator {\em fundamental} if its
levels of degeneracy are the lowest possible. It is natural to expect that
the fundamental singlet operator is the identity operator of the theory,
with dimension $\Delta_{\Phi^0} = 0$. Despite of this trivial value of
the scaling dimension, it is important to analyse the module of the identity
operator properly. Namely, the levels and $Z_N$ charges of its degenerate
submodules, and their multiplicities, provide important information that
we shall use in Section~3 to dress the general theory. This analysis was
done for $N=5$ in Ref.~\cite{para2}; we here extend it to $N=7$.

In the module of a singlet operator, shown in the top left part of
Fig.~\ref{fig1}, the first descendent level is the level $3/7$ in the $q=3$
sector. By using the commutation relations $\{\Psi^2,\Psi^1\}\Phi^0$ and
$\{\Psi^{-1},\Psi^{-3}\}\Phi^0$ (cf.~Eq.~(\ref{qdif})) it is easy to
verify that all the possible $q=3$ states at level $3/7$ are proportional
to one another
\beq
 A^1_{3/7} A^2_{-6/7} \Phi^0 \propto A^{-2}_{3/7} A^{-2}_{-6/7} \Phi^0 \propto
 A^2_{2/7} A^1_{-5/7} \Phi^0 \propto A^{-3}_{2/7} A^{-1}_{-5/7} \Phi^0 \propto
 A^{-1}_{0} A^{-3}_{-3/7} \Phi^0 \propto A^{3}_{-3/7}\Phi^{0}.
 \label{decompos}
\eeq
The above relations imply that there is only one $q=3$ state at level $3/7$.
Thus, in order to impose degeneracy at level $3/7$, we must require
that the state
\beq
 \chi^3_{-3/7} = A^3_{-3/7} \Phi^0 
\eeq
be primary. This amounts to the condition:
\beq
 A^{-3}_{+3/7} A^3_{-3/7} \Phi^0 = 0. 
 \label{constraint3_7}
\eeq

Using $\{\Psi^3,\Psi^{-3}\} \Phi^0$ (cf.~Eq.~(\ref{qmq})), it turns out that
the constraint (\ref{constraint3_7}) is equivalent to the following relation
between certain matrix elements at lower levels%
\footnote{Eq.~(\ref{constraint3_7_bis}) expresses just one of many equivalent
ways of transcribing the constraint (\ref{constraint3_7}) in terms of
relations between lower-lying matrix elements. Using the decompositions
(\ref{decompos}) it may be seen that among the various choices of rewriting
the constraint (\ref{constraint3_7}), the one that digs least deeply into
the module involves matrix elements up to level $6/7$. In any case, the key
point is that the matrix elements occuring in any such relation touch lower
levels than those that we degenerate in the present calculation.}:
\beq
 \left( A^3_{+24/7} A^{-3}_{-24/7}-\frac{27}{7}  
 A^3_{+17/7} A^{-3}_{-17/7}+\frac{270}{49} A^3_{+10/7}
 A^{-3}_{-10/7} \right) \Phi^0 =
 \left(1+\frac{48}{7c}\Delta_{\Phi^0}\right)\Phi^0.
 \label{constraint3_7_bis}
\eeq
Having imposed the constraint (\ref{constraint3_7})---or equivalently
the relation (\ref{constraint3_7_bis})---we can eliminate the only state
at level $3/7$. We thus set
\beq
 A^3_{-3/7} \Phi^0 = 0 . 
 \label{degeneracy3_7}
\eeq

We then consider the next available level $5/7$ in the $q=1$ charge sector. 
Once the level $3/7$ is empty, there are three main ways of descending to
this level:
\beq
 A^{1}_{-5/7}\Phi^0, \qquad A^{2}_{0} A^{-1}_{-5/7} \Phi^0,\qquad
 A^{-1}_{+1/7} A^{2}_{-6/7}\Phi^0.
 \label{states_5_7}
\eeq
However, among the above states  only two are independent. This is  
seen from the commutation relation $\{\Psi^{2},\Psi^{-1}\}\Phi^0$
from which it follows that
\beq
 A^{2}_{0} A^{-1}_{-5/7} \Phi^0 - A^{-1}_{+1/7} A^{2}_{-6/7}\Phi^0 =
 \frac{\lambda^{2,-1}_{1}}{6}  A^{1}_{-5/7}\Phi^0.
\eeq
In addition to the states (\ref{states_5_7}), we have to consider also the
states which are obtained by repeated application of the parafermionic zero
modes, like for example the state $A^{2}_{0}A^{-2}_{0} A^{2}_{0} A^{-1}_{-5/7}
\Phi^0$. By using the commutation relations (\ref{qdif}) and the condition
(\ref{degeneracy3_7}), we have checked that all these states actually
decompose into linear combinations of the states (\ref{states_5_7}). To see
this, it is sufficent to verify that
\beq
 A^{2}_{0} A^{-2}_{0} A^{1}_{-5/7} \Phi^0 = a  A^{1}_{-5/7}\Phi^0 +
 b A^{2}_{0} A^{-1}_{-5/7} \Phi^0,
 \label{decomposition}
\eeq
where $a$ and $b$ are some numerical coefficients.
  
Each state at level $5/7$ in the $q=1$ sector can thus be written in terms
of the states $A^{1}_{-5/7}\Phi^0$ and $A^{2}_{0} A^{-1}_{-5/7}\Phi^0$. Then,
if we demand the complete degeneracy at level $5/7$, we have to impose the
following conditions:
\bea
 A^{-1}_{+5/7} \left( A^{1}_{-5/7}\Phi^0 \right) &=& 0, \label{1condition} \\
 A^{1}_{+5/7} A^{-2}_{0} \left( A^{1}_{-5/7}\Phi^0 \right) &=& 0,
 \label{2condition} \\
 A^{-1}_{+5/7} \left( A^{2}_{0} A^{-1}_{-5/7}\Phi^0 \right) &=& 0,
 \label{3condition} \\  
 A^{1}_{+5/7}A^{-2}_{0} \left( A^{2}_{0} A^{-1}_{-5/7}\Phi^0 \right) &=& 0.
 \label{4condition} 
\eea
Using the charge conjugation symmetry, and Eq.~(\ref{decomposition}), it is
easy to see that if the conditions (\ref{1condition})--(\ref{2condition}) are
verified, the remaining equations (\ref{3condition})--(\ref{4condition})
are automatically satisfied. Once the conditions
(\ref{1condition})-(\ref{2condition}) have been imposed, all the states at
level $5/7$ can then be eliminated.

{}From the commutation relation $\{ \Psi^1,\Psi^{-1} \} \Phi^0$ we have that
\beq 
 A^{-1}_{+5/7}A^{1}_{-5/7}\Phi^0 = \frac{48}{7c} \Delta_{\Phi^0} \Phi^0.
\eeq
The condition (\ref{1condition}) then implies that
\beq
 \Delta_{\Phi^0} = 0.
\eeq
We have thus found the trivial scaling dimension of the identity.

The second equation (\ref{2condition}) simply determines the values of some
matrix elements, starting from the value of the matrix element $A^{1}_{+5/7}
A^{-2}_{0} A^{1}_{-5/7}\Phi^0$ which is set equal to zero. By using the
algebra of the parafermionic modes, other matrix elements can then be fixed.
For example, from the condition (\ref{1condition}) and the commutation
relations $\{\Psi^{-2},\Psi^{1}\}\Phi^0$ and $\{\Psi^{1},\Psi^{1}\}\Phi^2$, it
turns out that
\beq
 A^{1}_{+5/7}
 A^{-2}_{0} A^{1}_{-5/7}\Phi^0 = \lambda^{1,1}_{2}A^{-2}_{+6/7}
 A^{2}_{-6/7}\Phi^0 =0. 
\eeq

Once all the states at level $3/7$ and at level $5/7$ have been eliminated, it
can easily be shown that the level $6/7$ is automatically empty. In fact,
according to the commutation relation $\{\Psi^1\Psi^1\}\Phi^0$, the only
possible state $A^2_{-6/7}\Phi^0$ at level $6/7$ vanishes, since
\beq
 \lambda^{1,1}_{2}A^2_{-6/7}\Phi^0= A^1_{-1/7}A^1_{-5/7}\Phi^0 = 0.
\eeq

Summarising, we see that the fundamental singlet has one degenerate doublet 3
submodule at level $3/7$ and two degenerate doublet 1 submodules at level
$5/7$.

\subsubsection{Fundamental doublet 1 operator}

We here show how to extend the explicit degeneracy analysis of the fundamental
doublet 1 operator from the case $N=5$ (see Ref.~\cite{para2}) to $N=7$.

The first descendent states in this doublet (shown in the upper right
part of Fig.~\ref{fig1}) are the two conjugate doublet 2 states at
level $1/7$. We first impose complete degeneracy of these states; by
charge conjugation symmetry it suffices to study the one with $q=2$.
The degeneracy condition reads
\beq
 A^{-1}_{1/7} A^1_{-1/7} \Phi^1 = 0.
\eeq
By the commutation relations $\{\Psi^1,\Psi^{-1}\}\Phi^1$, this determines
a matrix element at level $2/7$:
\beq
 \mu_{1,-1} \Phi^1 \equiv A^1_{2/7} A^{-1}_{-2/7} \Phi^1 =
 \left( -\frac{6}{49}+\frac{24}{7c} \Delta_{\Phi^1} \right) \Phi^1,
\eeq
where $\Delta_{\Phi^1}$ is the conformal weight of the doublet
1 operator that were are trying to determine.

We then focus on the next available level, which is the singlet ($q=0$)
state at level $2/7$. Since there are no states left on level $1/7$,
our candidate for a singular state at level $2/7$ reads
\beq
 \chi^0_{-2/7} = a A^1_{-2/7} \Phi^{-1} + b A^{-1}_{-2/7} \Phi^1.
\eeq
The degeneracy conditions are $A^{\pm 1}_{2/7}\chi^0_{-2/7}=0$.
Using the charge conjugation symmetry, they can be summarised by
\beq
 \left( \mu_{1,1} \right)^2 = \left( \mu_{1,-1} \right)^2, \label{signambi}
\eeq
where the matrix element $\mu_{1,1}$ is defined by
\beq
 \mu_{1,1} \Phi^1 \equiv A^1_{2/7} A^1_{-2/7} \Phi^{-1} \label{mu11def}
\eeq
and $\mu_{1,-1}$ has been given above.
Using $\{\Psi^1,\Psi^1\} \Phi^{-1}$ we can evaluate
\beq
 \mu_{1,1} = \lambda^{1,1}_2 h_2,
\eeq
where the structure constant $\lambda^{1,1}_2$ is given by
Eq.~(\ref{ope1}), and the zero mode eigenvalue $h_2$ is
defined in Eq.~(\ref{zeroeigen}).

It would now appear natural to impose the third (and last) degeneracy
condition on the doublet 3 states at level $5/7$. But we shall now show
that this level is actually void, as a result of the degeneracies at
levels $1/7$ and $2/7$. Let us recall that there is by now zero states
at level $1/7$, and one state at level $2/7$ since we have set
\beq
 \chi^0_{-2/7} = A^1_{-2/7} \Phi^{-1} - A^{-1}_{-2/7} \Phi^1 \equiv 0.
\eeq
The potential states at level $5/7$ in, say, the charge $q=3$ sector
therefore read:
\beq
 A^2_{-5/7} \Phi^1, \qquad
 A^{-3}_{-5/7} \Phi^{-1}, \qquad
 A^3_{-3/7} A^{1}_{-2/7} \Phi^{-1}. \label{3states}
\eeq
The first of these states is zero as a consequence of the degeneracy
at level $1/7$, and of the commutation relations $\{\Psi^1,\Psi^1\} \Phi^1$
which imply
\beq
 A^1_{-4/7} A^1_{-1/7} \Phi^1 = \frac12 \lambda^{1,1}_2 A^2_{-5/7} \Phi^1.
\eeq
By charge conjugation we also have $A^{-2}_{-5/7} \Phi^{-1} = 0$. But the
commutation relations $\{\Psi^{-1},\Psi^{-2}\} \Phi^{-1}$ imply that
\beq
 A^{-1}_0 A^{-2}_{-5/7} \Phi^{-1} = \frac23 \lambda^{1,2}_3
 A^{-3}_{-5/7} \Phi^{-1},
\eeq
and we must therefore have $A^{-3}_{-5/7} \Phi^{-1} = 0$ as well.
Finally, it is a consequence of $\{\Psi^1,\Psi^3\} \Phi^{-1}$ that
\beq
 \left( A^1_{-4/7} A^3_{-1/7} - A^3_{-3/7} A^1_{-2/7} \right) \Phi^{-1} =
 -\frac14 \lambda^{1,3}_{-3} A^{-3}_{-5/7} \Phi^{-1}.
\eeq
We have shown that the first term on the left-hand side vanishes, as does
the term on the right-hand side. Therefore
$A^3_{-3/7} A^{1}_{-2/7} \Phi^{-1}$ must also vanish. In conclusion, we
have shown that all of the three states (\ref{3states}) are zero,
and thus level $5/7$ is empty.

We therefore try imposing degeneracy on the doublet 1 states at level 1.
For definiteness, we consider the charge sector $q=1$. There are
potentially three states at this position in the module:
\beq
 A^2_{-1} \Phi^{-1}, \qquad
 A^1_{-5/7} A^1_{-2/7} \Phi^{-1}, \qquad
 L_{-1} \Phi^1. \label{3etats}
\eeq
However, the first of these two states are dependent, as is seen by
using the commutation relations $\{\Psi^1,\Psi^1\} \Phi^{-1}$:
\beq
 A^1_{-5/7} A^1_{-2/7} \Phi^{-1} = \frac12 \lambda^{1,1}_2 A^2_{-1} \Phi^{-1}.
\eeq
It turns out to be most convenient to use this dependency to eliminate
the first of the states (\ref{3etats}). We therefore demand that
\beq
 \chi^1_{-1} = \tilde{a} L_{-1} \Phi^1 +
 \tilde{b} A^1_{-5/7} A^1_{-2/7} \Phi^{-1}
\eeq
be a singular state. Defining four matrix elements
\bea
 \mu_0 \Phi^1 &\equiv& L_1 L_{-1} \Phi^1, \label{matet1} \\
 \mu_1 \Phi^1 &\equiv& L_1 A^1_{-5/7} A^1_{-2/7} \Phi^{-1}, \\
 \mu_2 \left( A^1_{-2/7} \Phi^{-1} \right) &\equiv&
 A^{-1}_{5/7} L_{-1} \Phi^1, \label{matel3} \\
 \mu_3 \left( A^1_{-2/7} \Phi^{-1} \right) &\equiv&
 A^{-1}_{5/7} A^1_{-5/7} A^1_{-2/7} \Phi^{-1}, \label{matel4}
\eea
the degeneracy criterion can be cast in the form
\beq
 \mu_0 \mu_3 - \mu_1 \mu_2 = 0. \label{degcrit22}
\eeq

The first three of the matrix elements (\ref{matel1})--(\ref{matel3})
are easily found from the commutation relations
(\ref{virasoro})--(\ref{parvira}). The results are:
\beq
 \mu_0 = 2 \Delta_{\Phi^1}, \qquad
 \mu_1 = \frac{10}{7} \mu_{1,1}, \qquad
 \mu_2 = \frac{10}{7}, \label{mu012}
\eeq
where $\mu_{1,1}$ is defined by Eq.~(\ref{mu11def}). The evaluation of
the fourth matrix element is slightly more involved. Letting the commutation
relation $\{\Psi^1,\Psi^{-1}\} \Phi^0$ act on the state
$A^1_{-2/7} \Phi^{-1}$ we obtain
\beq
 \left( A^{-1}_{5/7} A^1_{-5/7} - \frac37 A^{-1}_{-2/7} A^1_{2/7}
 + A^1_{-2/7} A^{-1}_{2/7} \right) A^1_{-2/7} \Phi^{-1} =
 \frac{24}{7c} L_0 \left( A^1_{-2/7} \Phi^{-1} \right).
\eeq
This implies that
\beq
 \mu_3 = \frac37 \mu_{1,1} - \mu_{1,-1} + \frac{24}{7c}
 \left( \Delta_{\Phi^1} + \frac27 \right). \label{mu3}
\eeq

The solution of Eq.~(\ref{degcrit22}), with the ingredients
(\ref{mu012}) and (\ref{mu3}), depends on the sign chosen in
resolving Eq.~(\ref{signambi}). Choosing $\mu_{1,1} = +\mu_{1,-1}$,
we find the solutions
\beq
 \Delta^{(\pm)}_{\Phi^1} =
 \frac{1}{42} \left( 36 - c \pm \sqrt{(c-6)(c-216)} \right). 
 \label{doub1sol1}
\eeq
On the other hand, the choice $\mu_{1,1} = -\mu_{1,-1}$ leads to
\beq
 \Delta_{\Phi^1} = \frac{1}{84} \left( 128 + 5c \pm
 \sqrt{25 c^2 + 680 c + 16384} \right). 
 \label{doub1sol2}
\eeq
We shall see below that the solutions (\ref{doub1sol1}) are physically
acceptable, from the point of view of the general structure of the theory
that we are constructing, while (\ref{doub1sol2}) must be discarded as
non-physical. It is useful to rewrite Eq.~(\ref{doub1sol1})
in terms of the parametrisation (\ref{paramp}):
\bea
 \Delta_{\Phi^1}^{(+)}&=& \frac{5}{7} \frac{p+7}{p}, \label{dimfd1a} \\
 \Delta_{\Phi^1}^{(-)}&=& \frac{5}{7} \frac{p-5}{p+2}. \label{dimfd1b}
\eea

In summary, the fundamental doublet 1 operator has one degenerate doublet 2
submodule at level $1/7$, one degenerate singlet submodule at level $2/7$,
and one degenerate doublet 1 submodule at level $1$.

\subsubsection{Fundamental doublet $(N-1)/2$ operator}

Writing $N=2r+1$, the doublet with the largest possible $Z_{N}$ charge is the
doublet $r$. The fact that the module of the doublet $r$ has its two summits
in adjacent charge sectors is a help in the degeneracy computations. For this
reason, we have been able to treat the case of arbitrary odd $N$.

We begin by considering the case of $N=7$ in some detail. In this case,
the structure of the doublet $r=3$ is shown in the lower right part of
Fig.~\ref{fig1}. Using the commutation relations, it is easy to see
that all the ways of descending to level $2/7$ in the charge sector
$q=-1$ are in fact linearly dependent:
\beq
 A^3_{-2/7} \Phi^3 \propto A^1_{1/7} A^2_{-3/7} \Phi^3 \propto
 A^{-3}_{1/7} A^{-1}_{-3/7} \Phi^3 \propto A^{-2}_0 A^{-2}_{-2/7} \Phi^3
 \propto A^1_{1/7} A^1_{-3/7} \Phi^{-3} \propto A^2_{-2/7} \Phi^{-3}.
\eeq
In particular, there is only one $q=-1$ state at level $2/7$.
Imposing degeneracy of this state amounts to the condition
\beq
 A^{-2}_{2/7} A^2_{-2/7} \Phi^{-3} = 0. \label{deg27}
\eeq
This condition does not immediately lead to a fixation of the zero
mode eigenvalue $h_3$. Rather, it gives relations between certain matrix
elements deeper down in the module. The reason is essentially that
the commutation relations $\{\Psi^2,\Psi^{-2}\} \Phi^{-3}$ have a
high value of the parameter $\beta$, cf.~Eq.~(\ref{betadef}), and thus
several terms will contribute on the left-hand side. (We have already
encountered a similar phenomenon in Eq.~(\ref{constraint3_7_bis}).)

The next available level consists of the doublet 2 states at level $3/7$.
Due to the degeneracy (\ref{deg27}), level $2/7$ is now empty, and there
are a priori two states in the sector $q=-2$ at level $3/7$:
\beq
 A^2_{-3/7} \Phi^3 \quad \mbox{and} \quad A^1_{-3/7} \Phi^{-3}.
\eeq 
These states are however dependent, as is seen from
$\{\Psi^1,\Psi^1\} \Phi^{3}$. (Further candidate states obtained by acting
by the zero mode at level $3/7$ can be similarly eliminated.)
The degeneracy criterion reads
\beq
 A^{-1}_{3/7} A^1_{-3/7} \Phi^{-3} = 0. \label{deg37}
\eeq
By $\{\Psi^1,\Psi^{-1}\} \Phi^{-3}$, this immediately fixes the zero
mode eigenvalue:
\beq
 \left( h_3 \right)^2 = -\frac{5}{49} + \frac{24}{7c} \Delta_{\Phi^3}.
 \label{zeromoded3}
\eeq

As a consequence of $\{\Psi^1,\Psi^2\} \Phi^{-3}$ we have that
\beq
 \lambda_3^{1,2} A^3_{-4/7} \Phi^{-3} =
 A^1_{-2/7} A^2_{-2/7} \Phi^{-3} = -\frac32 A^2_{-1/7} A^1_{-3/7} \Phi^{-3}.
\eeq
Thus, since levels $2/7$ and $3/7$ are now empty, there are no states
left at level $4/7$.

With these simplifications, the next available states are in the $q=\pm 3$
sectors at level $1$. Focusing on $q=3$, the case of $q=-3$ being equivalent,
we form the candidate singular state
\beq
 \chi^3_{-1} = a L_{-1} \Phi^3 + b A^{-1}_{-1} \Phi^{-3}, \label{Z7chilev1}
\eeq
and we define the matrix elements
\bea
 \mu_{0,0}  \Phi^3    &=& L_1 L_{-1} \Phi^3, \\
 \mu_{0,-1} \Phi^3    &=& L_1 A^{-1}_{-1} \Phi^{-3}, \\
 \mu_{1,0}  \Phi^{-3} &=& A^1_1 L_{-1} \Phi^3, \\
 \mu_{1,-1} \Phi^{-3} &=& A^1_1 A^{-1}_{-1} \Phi^{-3}.
\eea
The degeneracy criterion then reads, as usual,
\beq
 \mu_{0,0} \mu_{1,-1} - \mu_{0,-1} \mu_{1,0} = 0. \label{degd3l1}
\eeq

The first three matrix elements are readily computed
\beq
 \mu_{0,0} = 2 \Delta_{\Phi^3}, \qquad
 \mu_{0,-1} = \mu_{1,0} = \frac{12}{7} h_3,
\eeq
and the fourth one is obtained from $\{\Psi^1,\Psi^{-1}\}\Phi^{-3}$,
yielding
\beq
 \mu_{1,-1} = \frac37 \left( h_3 \right)^2 + \frac{9}{49} +
 \frac{24}{7c} \Delta_{\Phi^3}.
\eeq

Inserting this into Eq.~(\ref{degd3l1}), and using also
Eq.~(\ref{zeromoded3}) for the zero mode eigenvalue, we arrive at the
solution
\beq
 \Delta^{(\pm)}_{\Phi^3} =
 \frac{1}{70} \left( 36 - c \pm \sqrt{(c-6)(c-216)} \right).
 \label{dimd3}
\eeq
This can also be written, using the parametrisation (\ref{paramp}), as:
\bea
 \Delta^{(+)}_{\Phi^3} &=& \frac37 \frac{p+7}{p}, \\
 \Delta^{(-)}_{\Phi^3} &=& \frac37 \frac{p-5}{p+2}.
\eea

In conclusion, the doublet 3 module that we have constructed has one
degenerate doublet 1 submodule at level $2/7$, one degenerate doublet 2
submodule at level $3/7$, and one degenerate doublet 3 submodule at level 1.

\underline{General case:} Based on the similarities of the fundamental
doublet $r=\frac{N-1}{2}$ modules for $N=5$ (see Ref.~\cite{para2}) and
$N=7$ (see above), we can conjecture the structure of this doublet in
the general case. One should demand the existence of $r$ distinct
degenerate submodules; however, only two degeneracy conditions are
required to fix the zero mode eigenvalue $h_r$ and the dimension
$\Delta_{\Phi^r}$. We conjecture that these two conditions are obtained
by requiring, respectively:
\begin{enumerate}
 \item The complete degeneracy of the doublet $r-1$ state
 at level $\delta^r_{r-1}=1-4/N$ (see Eq.~(\ref{levelall})).
 Note that we have here supposed that those of the remaining $r-2$ degeneracy
 conditions that act at levels strictly between $0$ and $\delta^r_{r-1}$ are
 such that these levels become empty after factoring out the corresponding
 degenerate submodules (or at least that no indirect ways of descending
 to level $\delta^r_{r-1}$ exist). Under this assumption, there remains
 only the two direct ways of descending to level $\delta^r_{r-1}$:
 \beq
  A^{-1}_{-\delta^r_{r-1}} \Phi^{r} \quad \mbox{and} \quad
  A^{-2}_{-\delta_{r-1}^{r}} \Phi^{-r}. \label{twostat}
 \eeq
 However, the commutation relation $\{\Psi^{-1},\Psi^{-1}\} \Phi^{-r}$
 of Eq.~(\ref{qdif}), with $n=1$ and $m=0$, shows that these two states
 are in fact proportional to one another.
 In order to have degeneracy at level $\delta_{r-1}^{r}$, it thus suffices
 to require that the state
 \beq
  \chi^{r-1}_{-\delta_{r-1}^{r}}\equiv A^{-1}_{-\delta_{r-1}^{r}}\Phi^{r}
  \label{stateD2}
 \eeq
 be primary. This condition then fixes the eigenvalue $h_r$ (see below).
 \item A degeneracy of the doublet $r$ state at level 1. We here suppose
 that those of the remaining $r-2$ degeneracy conditions that act at levels
 strictly between $\delta^r_{r-1}$ and 1 are such that these levels become
 empty after factoring out the corresponding degenerate submodules.
 If this is so, we can produce a degenerate state of a form
 analogous to Eq.~(\ref{Z7chilev1})
 \beq
  \chi^{r}_{-1}=a L_{-1}\Phi^{r}+bA^{-1}_{-1}\Phi^{-r}. \label{chidrlev1}
 \eeq
 Demanding the primarity of this state then fixes
 $\Delta_{\Phi^r}$ (see below). 
\end{enumerate}
Note that apart from eliminating unwanted states between levels 0 and 1,
the remaining $r-2$ degeneracy conditions may fix the values of certain
lower-lying matrix elements (as was the case for $N=7$, at level $2/7$).

We have verified that the scenario outlined above indeed holds
true also for $N=9$. Now, let us examine its algebraic consequences
in the general case.

The primarity of the state (\ref{stateD2}) can be obtained by requiring
that
\beq
 A^{1}_{+\delta_{r-1}^{r}}A^{-1}_{-\delta_{r-1}^{r}}\Phi^{r}=0.
\eeq
This can be rewritten, using the commutation relations
$\{\Psi^1,\Psi^{-1}\}\Phi^r$, through
\beq
 \left( A^{-1}_{0}A^{1}_{0}+A^{-1}_{\delta_{r-1}^{r}}
 A^{1}_{-\delta_{r-1}^{r}} \right) \Phi^{r} =
 \left( -\frac{N-2}{N^2}+\frac{2\Delta_{1}}{c}\Delta_{\Phi^{r}} \right)
 \Phi^{r},
\eeq
which fixes the eigenvalue (\ref{zeroeigen}) to be:
\beq
 \left( h_{r} \right)^2 =
 -\frac{N-2}{N^2}+\frac{2\Delta_{1}}{c}\Delta_{\Phi^{r}},
 \label{fixeig2}
\eeq
thus generalising Eq.~(\ref{zeromoded3}). If we impose
the condition (\ref{fixeig2}), then the state (\ref{stateD2}) can be set
equal to zero, thus reducing the module. With the above set of assumptions,
there are no states left in the module strictly between levels 0 and 1.

Turning now to level 1, we demand that the state (\ref{chidrlev1}) be
primary:
\beq
 L_{+1}\chi^{r}_{-1}=0,\quad A^{+1}_{+1}\chi^{r}_{-1}=0.
\eeq
In terms of the matrix elements $\mu_{ij}$ defined by
\bea
 L_{+1}L_{-1}\Phi^{r} &=&\mu_{11}\Phi^{r}, \\
 L_{+1}A_{-1}^{-1}\Phi^{-r} &=& \mu_{12}\Phi^{r}, \\
 A_{-1}^{-1}L_{-1}\Phi^{-r} &=& \mu_{21}\Phi^{r}, \\
 A_{+1}^{-1}A_{-1}^{1}\Phi^{r} &=& \mu_{22}\Phi^{r},
\eea
the degeneracy criterion reads
\beq
 \mu_{11}\mu_{22}-\mu_{12}\mu_{21}=0. \label{det0}
\eeq
Using the commutation relations, the required 
matrix elements are readily computed:
\bea
 \mu_{11} &=& 2\Delta_{\Phi^{r}}, \label{matel1}\\
 \mu_{12} = \mu_{21} &=& \Delta_{1}h_{r}, \\
 \mu_{22} &=& \frac{N-4}{N}(h_{r})^{2}+\frac{N+2}{N^2}+\frac{2\Delta_{1}}
 {c}\Delta_{\Phi^{r}},\label{matel2}
\eea
Inserting the eigenvalue (\ref{fixeig2}) and the matrix elements
(\ref{matel1})--(\ref{matel2}) into Eq.~(\ref{det0}) we get the following
solutions:
\beq
 \Delta^{(\pm)}_{\Phi^{r}} = \frac{1}{2}+\frac{1-c}{2N(N-2)} \pm
 \frac{1}{2N(N-2)}\sqrt{(c-N+1)(c-(N-1)^3)},\label{solD2ac}
\eeq
which generalise Eq.~(\ref{dimd3}).

Using the parametrisation (\ref{paramp}) 
in Eq.~(\ref{cparamv}), these solutions take the form
\bea
 \Delta^{(+)}_{\Phi^{r}}&=&\frac{1}{2}\frac{(N-1)(N+p)}{Np}, \label{solD2a} \\
 \Delta^{(-)}_{\Phi^{r}}&=&\frac{1}{2}\frac{(N-1)(p+2-N)}{(p+2)N}.
 \label{soluD2b}
\eea

It should be remarked that the assumptions that we have made in order to
determine the dimension $\Delta_{\Phi^r}$ for general $r$ will ultimately
be validated by the agreement of Eqs.~(\ref{solD2a})--(\ref{soluD2b}) with
the Kac formula which we shall discuss in the next Section.

\subsection{Sector of the disorder operators $ R_{a}$}
\subsubsection{Modes of the chiral operators and commutation relations}

The theory of disorder operators $R_{a}$ ($a=1,2,\ldots,N$) has been
fully developed in Ref.~\cite{dis1} within the first parafermionic
conformal field theory with symmetry $Z_N$, and in Refs.~\cite{para2_3,para2}
within the context of the second parafermionic theory with
symmetries $Z_3$ and $Z_5$ respectively. The general properties
(operator product expansions, analytic continuations, etc.) of the disorder
sector operators, and the approach used for studying this particular sector,
can be adapted to the present theory as well.

The non-abelian monodromy of the disorder operator
$R_{a}(z,\bar{z})$
 with respect to the chiral
 fields $\Psi^{\pm k}(z)$ amounts to the decomposition of
 the local products $\Psi^{k}(z) R_{a}(0)$
 into half-integer powers of $z$:
 \beq
 \Psi^{k}(z)R_{a}(0) =
 \sum_{n}\frac{1}{(z)^{\Delta_{k}+\frac{n}{2}}}\,
 A^{k}_{\frac{n}{2}}R_{a}(0), \qquad k=1,2,\ldots,\frac{N-1}{2}. \label{mode1R}
\eeq
The expansion of the product
$\Psi^{-k}(z)R_{a}(0)$ (with $k=1,2,\ldots,\frac{N-1}{2}$) can be obtained by
an analytic continuation of $z$ around 0 on both sides of
Eq.~(\ref{mode1R}). The result is:
\beq
 \Psi^{-k}(z)R_{a}(0) = \sum_{n}\frac{(-1)^{n}}{(z)^{\Delta_{k}+\frac{n}{2}}}
 A^{1}_{\frac{n}{2}} \, {\sf U}^k R_{a}(0), \qquad
 k=1,2,\ldots,\frac{N-1}{2}.
\eeq
In the above equation, ${\sf U}$ is a $N \times N$ matrix which
rotates the index of the disorder field: ${\sf U} R_{a}(0) =
R_{a-1}(0)$.

In accordance with these expansions, the mode operators
$A^{k}_{\frac{n}{2}}$ can be defined by the contour integrals
\beq
 A^{k}_{\frac{n}{2}}R_{a}(0)=\frac{1}{4\pi i}\oint_{C_{0}}
 {\rm d}z \, (z)^{\Delta_{k}+\frac{n}{2}-1}\Psi^{k}(z)R_{a}(0),
\eeq
where the integrations are defined by letting $z$ turn twice
around the operator $R_{a}(0)$ at the origin, exactly as
described in Ref.~\cite{para2_3}.

In the calculations of degeneracy we have used two types of
commutation relations: The first one is between the modes of
two $\Psi^1$ chiral fields,
\bea
 && \sum^{\infty}_{l=0} D^{l}_{\alpha\beta} \left(
 A^{1}_{\frac{n-l}{2}}A^{1}_{\frac{m+l}{2}} +
 A^{1}_{\frac{m-l}{2}}A^{1}_{\frac{n+l}{2}} \right) R_{a} \nn \\
 &=& \lambda^{1,1}_{2} 2^{\Delta_{2}-3}A^{2}_{\frac{n+m}{2}}R_{a}+
 (-1)^{n} 2^{-\Delta_{2}-2} \left[ \kappa(n)\delta_{n+m,0} +
 \frac{16\Delta_{1}}{c}
 L_{\frac{n+m}{2}} \right] {\sf U}^{-1}R_{a}, \label{psi1psi1}
\eea
and the second one is between the $\Psi^1$ and $\Psi^k$ chiral
fields, with $k=2,3,\ldots,\frac{N-1}{2}$,
\bea
 && \sum^{\infty}_{l=0} D^{l}_{\nu \mu} \left(
 A^{k}_{\frac{n-l}{2}}A^{1}_{\frac{m+l}{2}}-A^{1}_{\frac{m-l}{2}}
 A^{k}_{\frac{n+l}{2}} \right) R_{a} \label{psiqpsi1} \\
 &=& (-1)^{n+m} 2^{\mu-\nu-2}\lambda^{k,1}_{k+1}
 \left[(2\Delta_{k}+n-1)-\frac{\mu-\nu-2}{2}-a_{1}
 (2\Delta_{k+1}+n+m) \right]A^{k+1}_{\frac{n+m}{2}}R_{a} \nn \\
 &+& (-1)^{m} 2^{\nu-\mu-2}\lambda^{-k,1}_{1-k}
 \left[(2\Delta_{k}+n-1)-\frac{\nu-\mu-2}{2}-a_{2}(2\Delta_{1-k}+n+m)
 \right] A^{1-k}_{\frac{n+m}{2}} {\sf U}^{k} R_{a}, \nn 
\eea
where the coefficients $D^{l}_{\nu \mu}$ are defined by
\beq
 (1-x)^{\nu}(1+x)^{\mu}=\sum^{\infty}_{l=0}D^{l}_{\nu \mu}x^{l}.
\eeq
In Eq.~(\ref{psi1psi1}) we have used the following abbreviations:
\bea
 \alpha &=& 2\Delta_{1}-\Delta_{2}-1, \nn \\
 \beta  &=& 2\Delta_{1}-3, \nn \\
 \kappa(n)&=& (2\Delta_{1}+n-1)(2\Delta_{1}+n-2)-(2\Delta_{1}+n-1)
 (\Delta_{2}+1)+\frac{(\Delta_{2}+1)(\Delta_{2}+2)}{4}. \nn
\eea
And in Eq.~(\ref{psiqpsi1}) we have abbreviated the following quantities:
\bea
 \mu &=& \Delta_{k}+\Delta_{1}-\Delta_{k+1}-2, \nn \\
 \nu &=& \Delta_{-k}+\Delta_{1}-\Delta_{1-k}-2, \nn \\
 a_{1}&=&\frac{\Delta_{k+1}+\Delta_{k}-\Delta_{1}}{2\Delta_{k+1}}, \nn \\
 a_{2}&=&\frac{\Delta_{1-k}+\Delta_{k}-\Delta_{1}}{2\Delta_{1-k}}. \nn
\eea

\subsubsection{Degeneracy of the disorder modules}

The structure of the module of a disorder operator $R_a$ ($a=1,2,\ldots,N$) is
relatively simple, as witnessed by the expansion (\ref{mode1R}): Each module
has $N$ summits, labeled by the components $R_a$, and has only integer and
half-integer levels. Furthermore, it is seen from the expansion (\ref{mode1R})
that there are $(N-1)/2$ zero modes $A^{k}_{0}$ (with
$k=1,2,\ldots,\frac{N-1}{2}$), associated with the parafermion $\Psi^k$ which
acts between the $N$ summits of the module:
\beq
 A^{k}_{0}R_{a} = h_{k} \, {\sf U}^{2k}R_{a}. \label{Rzero1}
\eeq
This defines the eigenvalues $h_{k}$. We recall that ${\sf U} R_{a} = R_{a-1}$.
Like in the case of the doublet operators, the eigenvalues $h_{k}$
characterise, together with the conformal dimension, each representation
$R_a$. Actually, the eigenvalues $h_{1}$ and $h_{2}$ are linked by a relation
which does not depend on the details of the representation $R_a$. This
relation is easily obtained by setting $n=m=0$ in Eq.~(\ref{psi1psi1}):
\beq
 2h_1^2=\lambda^{1,1}_{2} 2^{\Delta_{2}-3}h_2+ 2^{-\Delta_{2}-2}
 \left[ \kappa(0)+\frac{16\Delta_{1}}{c} \Delta_R \right].
\eeq
We conclude then that each representation $R_a$ is characterised by
its dimension $\Delta_R$ and $(N-3)/2$ eigenvalues $h_{k}$.
These values are fixed by studying the degenerate representations of
the disorder sector. We shall show an example below.

The first descendants for a
given primary operator $R_{a}$ are found at level 1/2. For a given
value of the index $a$ there are $(N-1)/2$ states:
\beq
 \left( \chi^{(k)}_{a} \right)_{-\frac{1}{2}} =
 A^{k}_{-\frac{1}{2}} {\sf U}^{-2k}R_{a} =
 {\sf U}^{-2k} A^{k}_{-\frac{1}{2}}R_{a},
 \qquad k=1,2,\ldots,\frac{N-1}{2}. \label{Rstate1}
\eeq

In the case of $N=5$, there are two states,
$\left( \chi^{(1)}_{a} \right)_{-\frac{1}{2}}$ and
$\left( \chi^{(2)}_{a} \right)_{-\frac{1}{2}}$, in Eq.~(\ref{Rstate1}).
In Ref.~\cite{para2}, these were both required to be primary, and this choice
was shown to be consistent with the properties of the fundamental
disorder operator in a Coulomb gas construction based on the algebra $B_2$.
The most natural generalisation for $N>5$ is to impose that all of the states
(\ref{Rstate1}) be primary. The corresponding solutions give the dimensions of
the fundamental disorder operators, with the $(N-1)/2$ required degeneracies
all situated at the first descendant level $1/2$.

We therefore impose that
 \beq
 A^{k'}_{+\frac{1}{2}}(\chi^{(k)}_{a})_{-\frac{1}{2}}=0 \label{disdegcond}
\eeq
for each $k'=0,1,\ldots,\frac{N-1}{2}$, and for each
$k=0,1,\ldots,\frac{N-1}{2}$.  Using the commutation relations
(\ref{psi1psi1})--(\ref{psiqpsi1}), the degeneracy condition
(\ref{disdegcond}) results in a system of $(N-1)/2$ independent
equations which allow to determine the $(N-1)/2$ unknown variables,
i.e., the $(N-3)/2$ independent eigenvalues $h_{k}$ and the conformal
dimension $\Delta_{R}$ of the disorder operator $R_{a}$.  (We omit
here all the algebraic manipulations since this computation is
strictly analogous to the one presented in detail in
Ref.~\cite{para2}.)

Among the solutions for the conformal dimensions $\Delta_R$ admitted by this
system of $(N-1)/2$ equations, only two are physical solutions, in the sense
that they are consistent with the Kac formula which will be given in the next
Section. The two physical solutions for $\Delta_R$ read:
\beq
 \Delta^{(1)}_{R}  = \frac{1}{16}\frac{(N-1)(p+N)}{p}, \qquad
 \Delta^{(2)}_{R} = \frac{1}{16}\frac{(N-1)(p+2-N)}{p+2}. \label{soluR1}
\eeq

\section{Kac formula and boundary terms}

\subsection{Lie algebra structure}
\label{secLie}

It has already been observed in the Introduction that the central charge of the
second parafermionic theory with $Z_N$ symmetry agrees with that of a coset
based on the group $SO(N)$. It is therefore natural to suppose some connection
with the Lie algebra $B_r$, when $N=2r+1$ is odd. In the case $r=2$ this
connection was made explicit in Ref.~\cite{para2}, by assuming the existence
of a Coulomb gas realisation of the theory, based on the algebra $B_r$.
We shall here generalise this construction to arbitrary $r \ge 2$.

Each primary operator $\Phi$ of the theory is assumed to be represented by
a vertex operator, which can in turn be associated with the
weight lattice of $B_r$ as follows:
\beq
 \vec{\beta}\equiv\vec{\beta}_{(n_{1},\cdots,n_{r})(n'_{1},\cdots,n'_{r})}=
 \sum_{i=1}^{r} \left( \frac{1+n_{i}}{2}\alpha_{+} +
 \frac{1+n'_{i}}{2}\alpha_{-} \right) \vec{\omega}_{i}. \label{beta}
\eeq
Here, $\vec{\omega}_{i}$ are the fundamental weights of the algebra $B_{r}$,
and $\alpha_\pm$ are the usual Coulomb gas parameters. Their form
\beq
 \alpha_{+} = \sqrt{\frac{p+2}{p}},\qquad \alpha_{-}=-\sqrt{\frac{p}{p+2}}
 \label{alpha+-}
\eeq
are immediately suggested by Eq.~(\ref{cparamp}) for the central charge.

We shall use the notation
$\Phi_{(n_{1},\ldots,n_{r})(n'_{1},\ldots,n'_{r})}$ to represent
the operator associated with
$\vec{\beta}_{(n_{1},\ldots,n_{r})(n'_{1},\ldots,n'_{r})}$.
Note that the integers
$(n_{1},\ldots,n_{r})(n'_{1},\ldots,n'_{r})$ are essentially the Dynkin
labels of the weight $\vec{\beta}$. As usually in the Coulomb gas
construction the labels are doubled, with $(n_{1},\ldots,n_{r})$ representing
the $\alpha_+$ side of the theory, and $(n'_{1},\ldots,n'_{r})$ the
$\alpha_-$ side. The nature of a primary operator
$\Phi_{(n_{1},\ldots,n_{r})(n'_{1},\ldots,n'_{r})}$, i.e., its
transformation properties under $Z_N$, remains unchanged if the
$\alpha_+$ and $\alpha_-$ sides are interchanged.
Actually, such properties only depend on the differences $|n_i-n'_i|$.
In most of what follows we shall therefore choose to trivialise the
$\alpha_+$ side, by setting $(n_{1},\ldots,n_{r})=(1,\ldots,1)$.

The Coulomb gas formula for the conformal dimension of the primary operator
$\Phi_{(n_{1},\ldots,n_{r})(n'_{1},\ldots,n'_{r})}$ reads
\beq
 \Delta_{(n_{1},\cdots,n_{r})(n'_{1},\cdots,n'_{r})}\equiv
 \Delta_{(n_{1},\cdots,n_{r})(n'_{1},\cdots,n'_{r})}^{(0)}+B=
 \vec{\beta} \cdot \left( \vec{\beta}-2\vec{\alpha}_{0} \right)+B,
 \label{Kac}
\eeq
where the background charge $\vec{\alpha}_0$ is given by
\beq
 \vec{\alpha}_{0} = \left( \frac{\alpha_{+}+\alpha_{-}}{2} \right)
 \sum_{i=1}^{r} \vec{\omega}_{r}.
\eeq
In Eq.~(\ref{Kac}) we have written the scaling dimension as a sum
of two terms, $\Delta_{(n_{1},\ldots,n_{r})(n'_{1},\ldots,n'_{r})}^{(0)}$ and
$B$. The {\em Coulomb term}
$\Delta_{(n_{1},\ldots,n_{r})(n'_{1},\ldots,n'_{r})}^{(0)}$
depends on the parameters $\alpha^{2}_{+}$ and $\alpha^{2}_{-}$,
while the {\em boundary term} $B$ is a constant that characterises the sector
of the representation space to which the operator under consideration belongs.

Note that the identity operator is
$I = \Phi_{(1,\ldots,1)(1,\ldots,1)}$. Since
$\Delta_I = 0$ and $I$ is a singlet, we must necessarily have $B=0$ for
any singlet operator. However, we still need to find the correct values
of $B$ for the doublet $q$ and for the disorder sectors. These values,
and their assignment to the different positions of the weight lattice,
are arguably the main interest of the theory that we are constructing.

We shall refer to Eq.~(\ref{Kac}) as the Kac formula of the theory.
The table of operators $\Phi_{(n_{1},\ldots,n_{r})(n'_{1},\ldots,n'_{r})}$,
which is associated with the lattice made by the vectors
$\vec{\beta}_{(n_{1},\cdots,n_{r})(n'_{1},\cdots,n'_{r})}$,
shall be called the Kac table of the theory.

To use Eq.~(\ref{Kac}) to calculate actual values of the dimensions,
we need the
scalar products $\vec{\omega}_{i} \cdot \vec{\omega}_{j}$ for the Lie
algebra $B_r$, which are encoded in the quadratic form matrix:
\beq
 \vec{\omega}_{i} \cdot \vec{\omega}_{j}=i\,\,\mbox{for}\,\,i\leq j < r; \qquad
 \vec{\omega}_{i} \cdot \vec{\omega}_{r}=\frac{i}{2}\,\,\mbox{for}\,\,i<r;
 \qquad
 \vec{\omega}_{r} \cdot \vec{\omega}_{r}=\frac{r}{4}. \label{scalar}
\eeq

In what follows, a major role is played by the action of Weyl reflections
on the weights. We recall that the Weyl group $W$ is generated by $r$
{\em simple reflections} $s_{\vec{e}_i}$, with $i=1,2,\ldots,r$,
acting as follows on the vertex operators (weights):
\beq
 s_{\vec{e}_i} \vec{\beta}_{(1,\ldots,1)(n'_1,\ldots,n'_r)} =
 \vec{\beta}_{(1,\dots,1)(n'_1,\ldots,n'_r)} + n'_i \alpha_- \vec{e}_i.
\eeq
Here, $\vec{e}_i$ are the Lie algebra's simple roots,
whose coordinates in the basis of fundamental weights can be read off
from the rows of the Cartan matrix.
The total number of Weyl reflections is $|W|=2^r r!$. We also recall
that the Weyl group possesses a unique longest element, which can
be written as a word of length $r^2$ in terms of the generators
$s_{\vec{e}_i}$. This longest element simply changes the sign of
all the labels $(n'_1,\ldots,n'_r)$.

In the context of the conformal field theory,
the physical significance of the Weyl group is linked to the fact that the
simple roots represent the screenings operators of the Coulomb gas
realisation. As usual, we assume that the screenings commute
with the parafermionic algebra.

Given a generic vertex operator, there exists exactly one Weyl
reflection that maps it into the fundamental Weyl chamber,
the physical domain of the Kac table.
%
Like in Felder's resolution for minimal models \cite{Felder}, the
simple Weyl reflections indicate singular states (degeneracies) in
the modules of physical operators. More precisely, a simple
reflection can be associated with a mapping (realised by integrated
screenings) that takes a vertex operator outside the physical domain of the
Kac table into the module of a physical vertex operator. The difference between
the dimensions of these two operators, computed from Eq.~(\ref{Kac})
with the appropriate boundary terms, gives the level of degeneracy.
It is in the sense of this mapping that the
operator outside the physical domain is associated with a
non-physical (singular) state in the module of the physical operator; for this
reason it could be called a {\em ghost operator}.  The nature (singlet,
doublet $q$, disorder) of the corresponding singular state must
coincide with the labeling of the ghost operator. In a similar
fashion, non-simple reflections can be used to infer the fine
structure of the degenerate modules (ghosts within ghosts, etc.). For
instance, this aspect has been studied in a certain detail, in the case
of the representations of the $WA_2$ algebra,
in the paper \cite{Walg}, and references therein.

\subsection{Disorder sector}

We shall first consider the problem of the boundary term
for the disorder sector, and of the position of the fundamental
disorder operators.

The boundary term $B_{R}$ for this sector is readily accessible,
since we have computed explicitly the conformal dimensions of the
fundamental disorder operators (\ref{soluR1}) by means of
degeneracy conditions. Indeed, once we have identified the position
on the weight lattice of these operators by the coefficients of
$\alpha^{2}_{+}$ and $\alpha^{2}_{-}$, we may read off the boundary terms
from the $\alpha^{2}_{\pm}$ independent terms in the formulae for the
dimensions.

Having done some algebric manipulation, it is easy to verify---by
using the scalar products (\ref{scalar}) and Eq.~(\ref{beta})
in the Kac formula (\ref{Kac})---that the dimensions (\ref{soluR1}) of the
fundamental disorder operators correspond, respectively, to
\beq
 \Delta_{(1,\ldots,1,2)(1,\ldots,1)} \mbox{ and }
 \Delta_{(1,\ldots,1)(1,\ldots,1,2)}
 \label{rposi}
\eeq
with the boundary term
\beq
 B_{R}= \frac{N-1}{32}. \label{br}
\eeq

\subsection{Doublet sectors, $r=3$}

Identifying the boundary terms $B_{D^q}$ of the doublet $q$ sector is
more complicated. Indeed, we have not been able to explicitly compute
the dimensions of the fundamental doublet operators for general $r$.
However, in Section~2 we have obtained such explicit results for the
case of $r=3$ ($N=7$), at least for the doublet 1 and doublet 3 sectors.
Before turning to the general case, let us see how the $r=3$ case can
be completed by exploiting properties of the Coulomb gas representation.

Our explicit results for the dimensions of the fundamental doublet $1$
operators, see Eqs.~(\ref{dimfd1a})--(\ref{dimfd1b}), correspond respectively
to:
\beq
 \Delta_{(1,2,1)(1,1,1)} \mbox{ and }
 \Delta_{(1,1,1)(1,2,1)} \label{posfund1}
\eeq
with the boundary term
\beq
 B_{D^1}= \frac{3}{14}. \label{bd1}
\eeq
Simarly, the dimensions of the fundamental doublet $3$
operators, see Eqs.~(\ref{solD2a})--(\ref{soluD2b}), are given by:
\beq
 \Delta_{(2,1,1)(1,1,1)} \mbox{ and }
 \Delta_{(1,1,1)(2,1,1)},
\eeq
with the boundary term
\beq
 B_{D^3}= \frac{5}{28}. \label{bdr7}
\eeq

To determine the remaining boundary term $B_{D^2}$, for whose fundamental
operator we have no explicit degeneracy computation, we shall use an
argument based on Weyl reflections, as discussed in Section~\ref{secLie}.
To this end we consider the fundamental
doublet $1$ operator $\Phi_{(1,1,1)(1,2,1)}$, whose position was determined
in Eq.~(\ref{posfund1}) above. In this case the three principal reflections
are:
\bea
 s_{\vec{e}_1} \vec{\beta}_{(1,1,1)(-1,3,1)} =
 \vec{\beta}_{(1,1,1)(1,2,1)} &=& \vec{\beta}_{(1,1,1)(-1,3,1)} +
 \alpha_- \vec{e}_{1}, \nn\\
 s_{\vec{e}_2} \vec{\beta}_{(1,1,1)(3,-2,5)} =
 \vec{\beta}_{(1,1,1)(1,2,1)} &=& \vec{\beta}_{(1,1,1)(3,-2,5)} +
 2 \alpha_- \vec{e}_{2}, \nn\\
 s_{\vec{e}_3} \vec{\beta}_{(1,1,1)(1,3,-1)} =
 \vec{\beta}_{(1,1,1)(1,2,1)} &=& \vec{\beta}_{(1,1,1)(1,3,-1)} +
 \alpha_- \vec{e}_{3}. \label{dob2ref}
\eea
(In this expression, and in the following, the reflections are understood
to act on the $\alpha_-$ side of the indices only.)
These reflections give the following differences of the Coulomb part of
the dimensions:
\bea
 \Delta^{(0)}_{(1,1,1)(-1,3,1)}-\Delta^{(0)}_{(1,1,1)(1,2,1)}&=&\frac{1}{2},
 \label{baredelta0}\\
 \Delta^{(0)}_{(1,1,1)(3,-2,5)}-\Delta^{(0)}_{(1,1,1)(1,2,1)}&=&1, \\
 \Delta^{(0)}_{(1,1,1)(1,3,-1)}-\Delta^{(0)}_{(1,1,1)(1,2,1)}&=&\frac{1}{4}.
 \label{baredelta2}
\eea
In the above equations we have to add the correct boundary terms
in order to obtain the level of degeneracy on the right-hand side.
This amounts to determining to which sector the ghost operator belongs.
The important point is that the labeling of the ghost operator has to
coincide with the nature of the corresponding state in the module of the
doublet $\Phi_{(1,1,1)(1,2,1)}$. In Section~2 we have seen that the module of
the operator $\Phi_{(1,1,1)(1,2,1)}$ contains three singular states: A
doublet $2$ state at level $1/7$, a singlet state at level $2/7$, and a
doublet $1$ state at level $1$. It is then easy to verify,
using Eqs.~(\ref{bd1}), (\ref{bdr7}), and
(\ref{baredelta0})--(\ref{baredelta2}), that there is only one correct way
of labeling the ghost operators:
\bea
 \Phi_{(1,1,1)(-1,3,1)} &\sim& S,   \nn \\
 \Phi_{(1,1,1)(3,-2,5)} &\sim& D^1, \nn \\
 \Phi_{(1,1,1)(1,3,-1)} &\sim& D^2. \nn
\eea
The boundary term $B_{D^2}$ is then fixed by the condition
\beq
 \Delta^{(0)}_{(1,1,1)(1,3,-1)}+B_{D^2}-
 \Delta^{(0)}_{(1,1,1)(1,2,1)}-B_{D^1}=\frac{1}{7},
\eeq
which, using Eqs.~(\ref{baredelta2}) and (\ref{bd1}), yields:
\beq
 B_{D^2}=\frac{3}{28}. \label{bd2}
\eeq

Using the same kind of reasoning, we can also determine the nature of the
operator $\Phi_{(1,1,1)(1,1,3)}$. The principal reflections are:
\bea
 s_{\vec{e}_1} \vec{\beta}_{(1,1,1)(-1,2,3)} =
 \vec{\beta}_{(1,1,1)(1,1,3)} &=& \vec{\beta}_{(1,1,1)(-1,2,3)} +
 \alpha_- \vec{e}_{1} \nn\\
 s_{\vec{e}_2} \vec{\beta}_{(1,1,1)(2,-1,5)} =
 \vec{\beta}_{(1,1,1)(1,1,3)} &=& \vec{\beta}_{(1,1,1)(2,-1,5)} +
 \alpha_- \vec{e}_{2} \nn\\
 s_{\vec{e}_3} \vec{\beta}_{(1,1,1)(1,4,-3)} =
 \vec{\beta}_{(1,1,1)(1,1,3)} &=& \vec{\beta}_{(1,1,1)(1,4,-3)} +
 3 \alpha_- \vec{e}_{3}.
\eea
Examining the differences of dimensions as before, one easily checks that
the only acceptable configuration is:
\bea
 \Phi_{(1,1,1)(1,1,3)} &\sim& D^2, \\
 \Phi_{(1,1,1)(-1,2,3)} \sim \Phi_{(1,1,1)(2,-1,5)} &\sim& D^3, \nn \\
 \Phi_{(1,1,1)(1,4,-3)} &\sim& D^1. \nn
\eea
We infer from the above analysis that the module of the doublet $2$
operator $\Phi_{(1,1,1)(1,1,3)}$ has two singular doublet $3$ states at
level $4/7$ and one singular doublet $1$ state at level $6/7$.
Moverover, the operator $\Phi_{(1,1,1)(1,1,3)}$ must be the
{\em fundamental} doublet 2 operator, since these singular states are
situated at the lowest possible levels.

In summary, we have shown how to determine the boundary term $B_{D^2}$ by
exploiting the reflections on the weight lattice together with some partial
explicit results for the dimension of fundamental operators.

\subsection{Unitary theories. Finite Kac table}

All the considerations  done so far on the Kac table are valid for
general values of $\alpha_{+}$ and $\alpha_{-}=-1/\alpha_{+}$.
As usual, the Kac table becomes finite when
$\alpha^{2}_{+}= (p+2)/p$ takes rational values. The unitary
theories correspond to $p$ taking integer values, according to the
coset realisation (\ref{SOcoset})--(\ref{paramp}).

The finiteness of the Kac table can be shown by considering the reflections
with respect to the hyperplane defined by
\beq
 n'_{1}+2\sum_{i=2}^{r-1}n'_{i}+n'_{r}=p+2 \label{surface}
\eeq
on the $\alpha_-$ side, and by
\beq
 n_{1}+2\sum_{i=2}^{r-1}n_{i}+n_{r}=p \label{surface1}
\eeq
on the $\alpha_+$ side.
Geometrically, this hyperplane is perpendicular to the following
combination of screenings:
\beq
 \vec{e}_1+2\sum_{i=2}^{r} \vec{e}_i.
\eeq

Note that the differences between the dimensions of the operators linked by
such reflections are in agreement with the levels of the corresponding
modules. The operators above the hyperplane (\ref{surface}) are expected to
be ghost operators and decouple from the theory.
For $p$ integer, the physical part of the Kac table is therefore 
delimited by:
\bea
 2(r-1) &\leq& n'_{1}+2\sum_{i=2}^{r-1}n'_{i}+n'_{r}\leq p+1, \nn\\
 2(r-1) &\leq& n_{1}+2\sum_{i=2}^{r-1}n_{i}+n_{r}\leq p-1. \label{finite}
\eea

The finite Kac table defined by Eq.~(\ref{finite}) contains an extra symmetry.
It is easy to verify that the Coulomb part of the conformal dimension
$\Delta^{(0)}_{(n_{1},\ldots,n_{r})(n'_{1},\ldots,n'_{r})}$ in
Eq.~(\ref{Kac}), i.e., with the boundary term $B$ being neglected, is
invariant under the operation
\bea
 n'_{1}\rightarrow p+2-n'_{1}-2\sum_{i=2}^{r-1}n'_{i}-n'_{r},&&
 n'_{i}\rightarrow n'_{i} \quad i=2,3,\ldots,r \nn \\
 n_{1}\rightarrow p-n_{1}-2\sum_{i=2}^{r-1}n_{i}-n_{r},&&
 n_{i}\rightarrow n_{i} \quad i=2,3,\ldots,r.\label{symmetry}
\eea
Now, if the label of each operator (singlet, doublet $q$, or disorder)
is invariant under (\ref{symmetry}), the above reflection will stay a
symmetry when the boundary term is included. We shall see in Section~4
below that the correct labeling indeed has this property. Note that this
amounts to saying that the correct labeling has periodicity $2$ in the
$-\vec{\omega_{1}}/2$ direction.

\subsection{Values of the doublet boundary terms}

\subsubsection{Trivial theory: $c=0$}

Having defined the finite Kac tables, we can make an important observation.
{}From the formula (\ref{cparamv}), with the parametrisation (\ref{paramp}),
the central charge is equal to zero for $p=N-2$. (Note also that
the coset (\ref{SOcoset}) trivialises.) The corresponding theory is therefore
expected to be trivial, i.e., with all conformal weights equal to zero.

When $p=N-2$, it follows from Eq.~(\ref{finite}) that all physical operators
have trivial indices on the $\alpha_+$ side: $(n_1,\ldots,n_r)=(1,\ldots,1)$.
Moreover, using the symmetry (\ref{symmetry}) we see that the only allowed
excitations on the $\alpha_-$ side are the following:
\bea
 \Phi_{0} &\equiv& \Phi_{(1,\ldots,1)(1,1,\ldots,1,1,1,\ldots,1,1)},
 \nn \\
 \Phi_{k} &\equiv& \Phi_{(1,\ldots,1)(1,1,\ldots,1,2,1,\ldots,1,1)},
 \qquad 0<k<r,
 \nn \\
 \Phi_{r} &\equiv& \Phi_{(1,\ldots,1)(1,1,\ldots,1,1,1,\ldots,1,3)},
 \nn \\
 \Phi_{R} &\equiv& \Phi_{(1,\ldots,1)(1,1,\ldots,1,1,1,\ldots,1,2)},
 \label{trivial}
\eea
where in the second line we have $n'_k = 2$. As already discussed, the
operators $\Phi_{0}$ and $\Phi_{R}$ are, respectively, the identity operator
and a fundamental disorder operator $R$.
 
We assume that every physical operator in a $c=0$ theory should have
$\Delta=0$. When applied to the operators (\ref{trivial}), this allows
us to compute the values of all the boundary terms, for each
odd $N$, by using the Kac formula (\ref{Kac}). One obtains:
\bea
 B_R &=& \frac{N-1}{32} \nn \\
 B_k &=& \frac{k(N-2k)}{4N}, \qquad k=0,1,\ldots,\frac{N-1}{2}. 
 \label{Bterms}
\eea
In the above equation $B_R$ and $B_k$ denote, respectively, the boundary
terms of the operators $\Phi_{R}$ and $\Phi_{k}$, whose positions have
been defined in Eq.~(\ref{trivial}). The value of $B_R$ confirms
Eq.~(\ref{br})---the boundary term for a disorder operator---as it
should, since $\Phi_R$ is a fundamental disorder operator,
cf.~Eq.~(\ref{rposi}). Also, $\Phi_0$ is trivially a singlet and
corresponds to the boundary term $B_0 \equiv B_S = 0$.

The task of finding the boundary terms $B_{D^q}$ of the doublets $D^q$
has therefore been reduced to the problem of associating each value
$B_{k}$ to the correct charge sector or, in other words, to determine
the nature of the operators $\Phi_{k}$ with $k=1,2,\ldots,r$. We shall
assume that among the operators listed in Eq.~(\ref{trivial}), each
sector of the theory is represented exactly once. Since the singlet
and disorder sectors have already been accounted for, it follows that
the operators $\Phi_{k}$ with $k=1,2,\ldots,r$ are necessarily a
permutation of the (fundamental) doublet operators. Below, we shall
consider the problem of determining {\em which} permutation, for
general $r$.
 
\subsubsection{Position of the fundamental operators}

In Section~3.3 we have succeeded in determining the positions
on the weight lattice of the fundamental operators in each sector, in
the special case of $r=3$. The general structure of these results is most
easily seen if one introduces a new notation for the doublet sectors,
by doubling the $Z_N$ charges:
\beq
 Q = 2 q \mbox{ mod } N. \label{Qnotation}
\eeq
We shall distinguish between the two notations by placing an asterisk after
the indices which are to be read in the $Q$ notation. Note that since $N$
is odd, the $Q$ labels are simply a permutation of the $q$ labels.
The results of Section 3.3 can now be written in the form:
\bea
 \Phi_{(1,1,1)(2,1,1)} &\sim& D^3 \equiv D^{1^*}, \qquad \nn\\
 \Phi_{(1,1,1)(1,2,1)} &\sim& D^1 \equiv D^{2^*}, \qquad \nn \\
 \Phi_{(1,1,1)(1,1,3)} &\sim& D^2 \equiv D^{3^*}. \label{posfund7}
\eea
Comparing the positioning (\ref{posfund7}) with Eq.~(\ref{Bterms}), we
obtain the results (\ref{bd1}), (\ref{bdr7}), and (\ref{bd2}).

Moreover, we have explicitly computed the dimensions
of the fundamental doublet $D^{1^*}$ for each $r$ in Section~2.1.4.
By using the Kac formula (\ref{Kac}), it is easy to verify
that the dimensions (\ref{solD2a})--(\ref{soluD2b}) correspond,
respectively, to:
\beq
 \Delta_{(2,1,\ldots,1)(1,\ldots,1)} \mbox{ and }
 \Delta_{(1,\ldots,1)(2,1,\ldots,1)} 
\eeq
with the boundary term
\beq
 B_{D^{1^*}}= \frac{N-2}{4N}. 
 \label{Bd1*}
\eeq
Thus, $\Phi_{1}\sim D^{1^*}$,
and the value of the bondary term $B_{1}$ confirms (\ref{Bd1*}).

It is now not difficult to guess the labeling of the operators
$\Phi_{k}$ for general $r$:
\bea
 \Phi_0 \equiv \Phi_{(1,\ldots,1)(1,1,\ldots,1,1,1,\ldots,1,1)}
 &\sim& S=I, \nn \\
 \Phi_k \equiv \Phi_{(1,\ldots,1)(1,1,\ldots,1,2,1,\ldots,1,1)}
 &\sim& D^{k^*}, \qquad 0<k<r, \nn \\
 \Phi_r \equiv \Phi_{(1,\ldots,1)(1,1,\ldots,1,1,1,\ldots,1,3)} &\sim& D^{r^*}.
 \label{posconj}
\eea
Below we shall give a (partial) demonstration of (\ref{posconj}).

Consider a general physical operator $\Phi_{(1,\ldots,1)(n'_1,\ldots,n'_r)}$, 
for which the $r$ simple reflections $s_{\vec{e}_i}$ give:
\bea
 \Delta^{(0)}_{s_{\vec{e}_{i}}\vec{\beta}}-\Delta^{(0)}_{\vec{\beta}}&=&
 \frac{n'_{i}}{2}, \qquad i=1,2,\ldots,r-1, \nn \\
 \Delta^{(0)}_{s_{\vec{e}_{r}}\vec{\beta}}-\Delta^{(0)}_{\vec{\beta}}&=&
 \frac{n'_{r}}{4} \nn 
\eea
with $\vec{\beta}\equiv\vec{\beta}_{(1,\ldots,1)(n'_1,\ldots,n'_r)}$.
As discussed in Section~3.1, the ghost operator
$\Phi_{s_{\vec{e}_i}\vec{\beta}}$ (charge sector $q_2$) indicates a
singular state in the module of the physical operator
$\Phi_{\vec{\beta}}$ (charge sector $q_1$). Their respective boundary
terms $B_{D^{q_1}}$ and $B_{D^{q_2}}$ must then satisfy
\beq
 B_{D^{q_2}}-B_{D^{q_1}} + \frac{\tilde{n}}{4} = \delta^{q_1}_{q_2} + k_{1,2}. 
 \label{generalcon}
\eeq
Here, the gap $\delta^{q_1}_{q_2}$ is defined by Eq.~(\ref{levelall}),
$k_{1,2}$ is a non-negative integer, and $\tilde{n}$ is a positive
integer (which is even if $i<r$ above). The right-hand side of
Eq.~(\ref{generalcon}) is the degeneracy level of $\Phi_{\vec{\beta}}$
with respect to the singular state $\Phi_{s_{\vec{e}_i}\vec{\beta}}$.

The gaps in a generic singlet module are given by Eq.~(\ref{levelid}).
In the $Q$-notation this can be rewritten as
\beq
 \delta^{0}_Q = \frac{Q(N-Q)}{2N} \mbox{ mod } 1.
\eeq 
(Note that this is not obtained naively, by substituting $Q=2q$
in Eq.~(\ref{Qnotation}).) Thus, choosing in particular the physical
operator to be a singlet, $q_1=0$ and $B_0=0$, the condition
(\ref{generalcon}) fixes the general form of the boundary term $B_{D^{Q}}$:
\beq
 B_{D^{Q}}= \frac{Q(N-Q)}{2N}-\frac{k_{Q}}{4},
 \label{Bgeneralform}
\eeq
where $k_{Q}$ is an integer which depends on $Q$. Note that the
set of values $k_{Q}$ can be fixed, for example, from knowledge of
the levels of degeneracy of the fundamental operators.   

If $N$ is a prime number it is easy to prove, by comparison of
Eqs.~(\ref{Bgeneralform}) and (\ref{Bterms}), that $k_{Q}=Q$ and so
\beq
 B_{D^{q*}}=B_{q*}, \qquad q*=0,1,\ldots,r. 
 \label{solutiont}
\eeq
This then proves Eq.~(\ref{posconj}). 

If $N$ is not prime, some ambiguity remains. Let us consider the quantity
\beq
 \delta_{Q'}^0 - \delta_{Q''}^0 =
 \left( \frac{Q'-Q''}{2} - \frac{(Q'+Q'')(Q'-Q'')}{2N} \right) \mbox{ mod } 1.
\eeq
Note that $|Q' \pm Q''| < N$, and so if $N$ is prime the right-hand
side can only vanish if $Q'=Q''$. On the other hand, if $N=s t$ for
some integers $s,t > 1$, the right-hand side can vanish for $Q' \neq
Q''$ provided that $|Q'-Q''| \mbox{ mod } s = 0$ and $|Q'+Q''| \mbox{
mod } t = 0$ (or vice versa).  In this case we would have $\delta_{Q'}^0
= \delta_{Q''}^0$, and the argument leading to (\ref{solutiont})---which
was based on the level structure of the modules---would break down; we
would not be able to distinguish between the charge sectors $Q'$ and
$Q''$. In particular, in addition to (\ref{solutiont}) the solution
\bea 
 B_{D^Q} &=& B_{Q}, \qquad Q=0,1,\ldots,r; \quad Q \neq Q', Q'', \nn \\
 B_{D^{Q'}}&=&B_{Q''}, \label{solutionf} \\
 B_{D^{Q''}} &=& B_{Q'} \nn
\eea
would a priori be acceptable.%
\footnote{For $N=s^2$, one would have $Q'=0$ and $Q''=s$. In this case there
is no ambiguity as the boundary term of the singlet sector is trivially
$B_0 = 0$. The positioning (\ref{posconj}) of the fundamental operators 
is then the only one that is acceptable.}

Summarising, we have explicitly proved for $r\leq 6$ that the
positioning (\ref{posconj}) is the only correct one. For $r>6$  
we have no strict arguments to exclude some additional
possibilities, such as (\ref{solutionf}). However, we find it natural to
assume that the theories that we are constructing should have a similar
structure for all $r$. We shall therefore accept the solution
(\ref{solutiont}) for all $r \ge 1$. 

Summarising, we write here the final result for the boundary terms:
\bea
 B_R       &=& \frac{N-1}{32}, \nn \\
 B_S       &=& 0, \nn \\
 B_{D^{Q}} &=& \frac{Q(N-2 Q)}{4N}, \qquad Q=1,2,\cdots,\frac{N-1}{2}. 
 \label{Btermsf}
\eea

\subsection{Alternative derivation of doublet boundary terms}

We now briefly show that it is possible to derive the boundary terms
of the doublet sectors without making reference to the $c=0$ theories.
In this second argument, which is based on the technique of Weyl
reflections, we make two important assumptions:
\begin{itemize}
 \item The positions of the fundamental operators are given by (\ref{posconj}).
 \item The levels of degeneracy $\delta$ of a fundamental operator belong to
  the interval $0 < \delta \le 1$.
\end{itemize}

We start by applying the $r$ simple Weyl reflections $s_{\vec{e}_i}$ to
the identity operator $I=\Phi_{(1,\ldots,1)(1,\ldots,1)}$. This yields the
following differences of dimensions:
\bea
 \Delta^{(0)}_{(1,\ldots,1)s_{\vec{e}_i}(1,\ldots,1)}
 -\Delta_{(1,\ldots,1)(1,\ldots,1)}
 &=&\frac{1}{2}, \qquad i=1,2,\ldots,r-1, \label{difidi} \\
 \Delta^{(0)}_{(1,\ldots,1)s_{\vec{e}_r}(1,\ldots,1)}
 -\Delta_{(1,\ldots,1)(1,\ldots,1)}&=&\frac{1}{4}, \label{difidr}
\eea
with $\Delta_{(1,\ldots,1)(1,\ldots,1)}=0$.
The only labeling consistent with Eqs.~(\ref{difidi})--(\ref{difidr}) is
\bea
 \Phi_{(1,\ldots,1)s_{\vec{e}_i}(1,\ldots,1)} &\sim& D^1, \qquad
 i=1,2,\ldots,r-1, \nn \\
 \Phi_{(1,\ldots,1)s_{\vec{e}_r}(1,\ldots,1)} &\sim& D^r,
\eea
and in addition we deduce the boundary term $B_{D^1}=\frac{N-4}{2N}$. Note
that this implies that the module of the identity has a singular doublet
$r$ state at level $(N-1)/2N$ and $r-1$ singular doublet $1$ states at level
$(N-2)/N$. In the special case of $N=7$, our explicit degeneracy computations
have indicated the mechanism by which the doublet $1$ states may proliferate,
as required by this scenario. In particular, we expect that for $N>7$ the
relation analogous to (\ref{decomposition}) no longer holds true.

With the boundary term $B_{D^1}$ in hand, we next consider the simple
Weyl reflections of the fundamental doublet $1$ operator
$\Phi_{(1,\ldots,1)(1,2,1,\ldots,1)}$. In particular, the reflection
$s_{\vec{e}_r}$ gives a doublet $r$ singular state which, according
to Eq.~(\ref{levelall}), should be found on level $\delta^1_{r-1}=(N-5)/2N$.
This leads to the identity
\beq
 \Delta^{(0)}_{(1,\ldots,1)(1,2,1,\ldots,1,2,-1)}+B_{D^{r-1}}-
 \Delta^{(0)}_{(1,\ldots,1)(1,2,1,\ldots,1)}-B_{D^1}
 =\frac{N-5}{2N},
\eeq
from which we determine the boundary term $B_{D^{r-1}} = \frac{3(N-6)}{4N}$.

Knowing now the corresponding boundary term, we can consider the reflections
of the fundamental doublet $r-1$ operator, etc. Proceeding like this in a
systematic way, one can determine all boundary terms (\ref{Btermsf}).

\section{General theory}

Having fixed the positions of the fundamental operators and determined the
values of all the boundary terms of the theory, the remaining problem is to
fill the rest of the Kac table. In other words, we have to assign correctly
the boundary terms (\ref{Bterms}) to the Kac formula (\ref{Kac}) for every
vertex of the weight lattice. This amounts to determining the sector
label with respect to the group $Z_N$ of each operator in the theory
(the case of the fundamental operators was settled in Section~3.5).

After having addressed this question, we dedicate the rest of the present
Section to a study of the characteristic equations of various three-point
functions. These will give us independent verifications of a number of
aspects of our theory, and also yield some new results.

\subsection{Filling the lattice}
\label{fillingthelattice}

Although the Kac table of the $Z_N$ theory, with $N=2r+1$, is based
on a $2r$-dimensional lattice, it suffices to fill one ``basic layer'', for
instance on the $\alpha_{-}$ side, whose vertices have the positions
$\vec{\beta}_{(1,\ldots,1)(n'_{1},\ldots,n'_{r})}$. The filling of the other
layers is obtained by shifting the filling of the basic layer. Namely,
the nature of the operator $\Phi_{(n_{1},\ldots,n_{r})(n'_{1},\ldots,n'_{r})}$
depends only on the differences $|n_{i}-n'_{i}|$.

Until now we have extensively used the fact that the degeneracies in the
modules could be read off from the Weyl reflections of the weight lattice. In
this section we shall exploit another useful rule of the Coulomb gas
representation.

We consider the multiplication (fusion) of two operators in the basic layer,
\beq
 \Phi_{(1,\ldots,1)(n'_{1},\ldots,n'_{r})} \cdot
 \Phi_{(1,\ldots,1)(m'_{1},\ldots,m'_{r})}.
\eeq
According to the Coulomb gas rules this produces, in the principal channel,
an operator $\Phi_{(1,\cdots,1)(l'_{1},\cdots,l'_{r})}$ with
\bea
 \vec{\beta}_{(1,\ldots,1)(l'_{1},\ldots,l'_{r})}&=&
 \vec{\beta}_{(1,\ldots,1)(n'_{1},\ldots,n'_{r})}
 +\vec{\beta}_{(1,\ldots,1)(m'_{1},\ldots,m'_{r})} \nn \\
 l'_{i}&=&n'_{i}+ m'_{i}-1, \qquad i=1,2,\ldots,r. \label{fusionrule}
\eea
The non-principal channels follow the principal one by shifts realised
by linear combinations (over the integers) of the simple roots $\vec{e}_{i}$
(with $i=1,\ldots,r$).

The above rule is extremely useful in fixing the distribution of operators on
the lattice. It should however be used with some care, since the amplitude of
the principal channel may vanish (see below).

\subsubsection{Disorder operators}

Postponing for the moment the potential difficulty of vanishing
amplitudes, we begin by examining for example the fusion of the
fundamental disorder operator $\Phi_{(1,\ldots,1)(1,\ldots,1,2)}$ and the
fundamental doublet $r$ operator $\Phi_{(1,\ldots,1)(2,1,\ldots,1)}$.
The result must necessarily be another disorder operator, since in the dihedral
group the product of a reflection and a rotation yields another reflection.
Thus, $\Phi_{(1,\ldots,1)(2,1,\ldots,1,2)}$ is
a disorder operator. Proceeding in this way it is easy to verify that the
operator $\Phi_{(1,\ldots,1)(n'_{1},\ldots,n'_{r})}$ with $|n'_{r}-1|$ odd is
a disorder operator $R$. More generally:
\beq
 \Phi_{(n_{1},\ldots,n_{r})(n'_{1},\ldots,n'_{r})} \sim R, \mbox{ when }
 |n_{r}-n'_{r}| \mbox{ is odd}. \label{Rpositions}
\eeq
We have thus filled half of the lattice with disorder operators.

An independent verification of this result can be obtained by the method of
reflections. First, it is not difficult to see that all Weyl reflections,
applied to an arbitrary operator $\Phi_{(1,\ldots,1)(n'_1,\ldots,n'_r)}$
in the basic layer, conserve the parity of $n'_{r}$. This is at least
consistent with the above separation between disorder and singlet/doublet
operators.

To perform a more detailed check, we consider the $r$ simple Weyl reflections
of the fundamental disorder operators $\Phi_{(1,\ldots,1)(1,\ldots,1,2)}$.
These map the ghost operator
$\Phi_{(1,\ldots,1)(m'^{(i)}_{1},\ldots,m'^{(i)}_{r})}$, with
$i=1,2,\ldots,r$, into the module of the disorder operator
$\Phi_{(1,\ldots,1)(1,\ldots,1,2)}$. The indices of the ghost operator read
\bea
 m'^{(i)}_{j}  &=& 1, \mbox{ for } j<i-1 \mbox{ and } i+1<j<r; \nn \\
 m'^{(i)}_{i-1}&=& m^{(i)}_{i+1}=2; \qquad m'^{(i)}_{i}=-1; \qquad
 m'^{(i)}_{r}=2 \nn
\eea
for $i<r$, while for $i=r$ they read
\beq
 m'^{(r)}_{j}=1, \mbox{ for } j<r-1; \qquad
 m'^{(r)}_{r-1}=3; \qquad m^{(r)}_{r}=-2.
\eeq
All the $r$ simple reflections give the result $1/2$ for
the difference of dimensions
\beq
 \Delta^{(0)}_{(1,\ldots,1)(m'^{(i)}_{1},\ldots,m'^{(i)}_{r})}-
 \Delta^{(0)}_{(1,\ldots,1)(1,\ldots,1,2)}= \frac{1}{2},
 \mbox{ for } i=1,2,\ldots,r. \label{difdis}
\eeq
This on one hand confirms the fact that the fundamental disorder operator
$\Phi_{(1,\ldots,1)(1,\ldots,1,2)}$ has $r$ singular states at level $1/2$,
as was supposed in Section~3. On the other hand, it classifies all the
concerned ghost operators
$\Phi_{(1,\ldots,1)(m'^{(i)}_{1},\ldots,m'^{(i)}_{r})}$
as disorder operators.

More generally, it is not difficult to see that the difference between
the Coulomb term of the dimension of any operator
$\Phi_{(1,\ldots,1)(n'_1,\ldots,n'_r)}$ with even $n'_r$ and the ghost
operators obtained by acting on it with (not necessarily simple)
reflections equals a half-integer. This is in strong support of the
result (\ref{Rpositions}).

\subsubsection{Singlet operators. Periodicity of the filling}

It remains to assign the doublet $D^q$ operators and the singlet operators
$S$ to the lattice sites $\Phi_{(1,\ldots,1)(n'_1,\ldots,n'_r)}$ with
$n'_r$ odd.

The singlet operator $S$ form a subalgebra, as the fusion between two
singlets $S \cdot S$ produces another singlet. Furthermore, the fusion
$D^q \cdot S$ gives another doublet $D^q$. The importance of the above
consideration is that once the positions of the first non-trivial
singlets (different from the identity) have been found, along each of
the principal lattice directions, these will set the periodicity of
the labeling of all lattice sites.  To see this, one first considers
the fusion of these non-trivial singlets among themselves, the result
being a periodic distribution of singlet operators throughout the
lattice. The periodicity in each lattice direction is given by the
distance between the first non-trivial singlet in that direction and
the identity operator. The remaining (non-singlet) labels can now be
translated throughout the lattice by fusing the corresponding
operators with all available singlets.

As noticed above, it might happen that the amplitude of the principal channel
vanishes. But in view of the role of the singlets in setting the periodicity
of the filling, it seems natural to suppose that this does not happen in
fusions involving singlets.

Notice that the symmetry (\ref{symmetry}) of the finite Kac tables implies
that the periodicity in the $n'_1$ direction must be $2$. The remaining
part of the task is to find the periodicities in the other lattice directions,
and to specify the labeling of the sites in a $r$-dimensional hypercube
of the lattice that corresponds to the given periodicity.

\subsubsection{Doublet operators}

To determine this periodicity, and to position the required non-fundamental
doublet operators, we have first tried using the method of fusions.
Consider for instance fusing the fundamental doublet $D^{1*}$ with itself
\beq
 \Phi_{(1,\ldots,1)(2,1,\ldots,1)} \cdot
 \Phi_{(1,\ldots,1)(2,1,\ldots,1)} \sim
 \Phi_{(1,\ldots,1)(3,1,\ldots,1)}.
\eeq
The result can either be a doublet $D^{2*}$ or a singlet $S$, since the
addition of $Q$-charges $(\pm 1) + (\pm 1)$ is ambiguous. Further fusions can
be used to establish that there is only a finite number of possible
periodicities along each lattice direction (in particular, the periodicity
cannot exceed $r$), but no definite answer emerges. Moreover, in the
$Z_{2r+1}$ theory with $r=3$, which we have studied in great detail, the
application of the fusion method leads to a number of inconsistencies with
the method of Weyl reflections. We conclude that the fusion method should
be abandoned, since the fusion of two particular doublets can (and does;
see below) lead to vanishing amplitudes in a few cases.

We therefore turn to the method of reflections. The idea is that the
number of elements in the complete Weyl group, $|W|=2^r r!$, is in
general very large. Thus, classifying $|W|-1$ ghost operators for each
fundamental operator---whose positions are supposed to be given by
Eq.~(\ref{posconj})---will fix the labels of a large number of
lattice sites close to the origin, hopefully enabling us to discern
the correct periodicity.

We have pursued this idea for $r=3$ and $r=4$, using a computer.
As a first step, the Weyl group was generated in a recursive fashion, as
follows. Recall that the elements of the Weyl group can be represented as
words made out of an $r$-letter alphabet,
$s_{\vec{e}_1},s_{\vec{e}_2},\ldots,s_{\vec{e}_r}$, each word having
it minimal possible length. In the case of the algebra $B_r$, there is
a unique longest element represented by a word of length $\ell_0 = r^2$;
moreover, all lengths $\ell$ satisfying $1 \le \ell \le \ell_0$ correspond
to at least one element. Therefore, all elements of length $\ell$ can be
generated by prefixing the words representing elements of length $\ell-1$
by each of the $r$ letters in the alphabet. For each word generated, one
tests whether it represents a new element of the Weyl group by considering
its action on a fixed lattice site in the fundamental chamber, and comparing
to the action of elements represented by shorter words. The recursion starts
from the simple Weyl reflections, and terminates once an element of length
$\ell_0$ has been generated.

For $r=3$, the resulting classification of $(r+1)|W| = 192$ singlet and
doublet operators made it evident that the correct filling has periodicity
2 in the $n'_i$ direction, for $i=1,2,\ldots,r-1$, and periodicity 4 in
the $n'_r$ direction. In particular, the first non-trivial singlets
(different from the identity) along each lattice direction are situated at
\bea
  \Phi_{(1,\ldots,1)(1,1,\ldots,1,3,1,\ldots,1,1)} &\sim& S, \qquad 0<k<r,
  \nn \\
  \Phi_{(1,\ldots,1)(1,1,\ldots,1,1,1,\ldots,1,5)} &\sim& S,
  \label{periodicity}
\eea
where in the first line $n'_k = 3$. Fusions by these singlets subsequently
permitted us to label all operators in the hypercubic unit cell
\bea
  & &1 \le n'_k \le 2, \qquad 0 < k < r, \nn \\
  & &1 \le n'_r \le 4, \label{unitcell}
\eea
whose translation throughout the lattice yields the complete filling.
Note that the sites in the cell (\ref{unitcell}) with $n'_r=2,4$ have
already been classified as disorder operators, by Eq.~(\ref{Rpositions}).

The results (\ref{periodicity})--(\ref{unitcell}) on the periodicity
was confirmed by the case $r=4$, where $(r+1)|W| = 1920$ singlet and
doublet operators on the four-dimensional weight lattice were classified.%
\footnote{It is worth mentioning a slight problem that appears in the analysis.
Namely, for $r=4$ the reflection method cannot distinguish between $S$
and $D^2$ operators. This is essentially because the Coulomb part of the
dimensions (\ref{Kac}) of two operators linked by a Weyl reflection
always differ by $p/4$, where $p$ is an integer. It is this difference,
taken modulo 1, that serves to classify a given ghost operator.
With $r+1$ distinct singlet and doublet sectors, this will necessarily
lead to ambiguities for $r \ge 4$. (In the case $r=4$ the ambiguity can
be resolved by using the known positions of the fundamental operators.)
These ambiguities are related to
those discussed near Eq.~(\ref{solutionf}).}

We now give the general result on the labeling of the unit cell
(\ref{unitcell}). It is most easily stated in terms of the
$Q$-notation for the doublets, cf.~Eq.~(\ref{Qnotation}). Defining:
\bea
 x_{i}&=& |n_{i}-n'_{i}| \mbox{ mod } 2, \qquad i=1,2,\ldots,r-1, \nn \\
 x_{r}&=& \frac{|n_{r}-n'_{r}|}{2} \mbox{ mod } 2,
\eea
the doublet charge $Q$ associated with the position
$\vec{\beta}_{(n_{1},\ldots,n_{r})(n'_{1},\ldots,n'_{r})}$ is given by
the recursive formula
\beq
 Q(x_1,x_2,\ldots,x_{k-1},1,0,\ldots,0) =
 k -  Q(x_1,x_2,\ldots,x_{k-1},0,0,\ldots,0), \label{filling}
\eeq
with the initial condition $Q(0,\ldots,0) = 0$. For the case of $r=3$
this reads explicitly:
\bea
 \Phi_{(1,1,1)(1,1,1)} \sim S, \ \ \ \quad
 \Phi_{(1,1,1)(1,2,1)} \sim D^{2^*}, \quad
 \Phi_{(1,1,1)(1,1,3)} \sim D^{3^*}, \quad
 \Phi_{(1,1,1)(1,2,3)} \sim D^{1^*}, \nn \\
 \Phi_{(1,1,1)(2,1,1)} \sim D^{1^*}, \quad
 \Phi_{(1,1,1)(2,2,1)} \sim D^{1^*}, \quad
 \Phi_{(1,1,1)(2,1,3)} \sim D^{2^*}, \quad
 \Phi_{(1,1,1)(2,2,3)} \sim D^{2^*}. \nn
\eea

It is easy to prove from Eq.~(\ref{filling})---using an induction
argument---that the number of doublet $D^Q$ operators
in the unit cell (\ref{unitcell}) equals ${r \choose Q}$, for
$Q=0,1,\ldots,r$.

Note that the filling (\ref{filling}) implies that some of the
doublet-doublet fusions must have a vanishing amplitude in the
principal channel. For $r=3$, an example is
\beq
 \Phi_{(1,1,1)(1,2,1)} \cdot \Phi_{(1,1,1)(2,1,3)} \to
 \Phi_{(1,1,1)(2,2,3)}.
\eeq
According to Eq.~(\ref{filling}), both operators on the left-hand
side are $D^{2*}$. Using the Coulomb gas fusion rules, the resulting
operator should thus be either a $S$ or a $D^{3*}$.
However, the operator $\Phi_{(1,1,1)(2,2,3)}$ appearing on the
right-hand side has been classified as $D^{2*}$ in Eq.~(\ref{filling}).
Therefore the amplitude of the principal channel must vanish.


\subsection{Characteristic equations}

We now present some additional verifications of the theory we have
constructed. To this end, we consider the characteristic equations for the
conformal dimensions of the operators which participate in a given fusion
(operator product expansion). These equations can be derived from the study of
the three-point correlation functions in a way analogous to that of
Refs.~\cite{para2_3}.

The three-point correlation functions considered below have the same
analytical properties as the ones studied in Ref.~\cite{para2} for $N=5$.
Therefore, for the more general case $N=2r+1$, the derivation of the
corresponding characteristic equations is the same. The details of the
calculations can be found in the Appendix C of Ref.~\cite{para2}.

We have considered the following correlation functions:
\begin{itemize}
\item
$\left \langle R^{(2)}(z_{2})\Phi^{q*}(z_{3})R^{(1)}(z_{1}) \right \rangle$

Here, $R^{(1)}(z_{1})$ is a disorder operator whose module is
supposed to be completely degenerate at level 1/2, i.e.,
$R^{(1)}(z_{1})=\Phi_{(1,\ldots,1,2)(1,\ldots,1)} (z_{1})$ or
$\Phi_{(1,\ldots,1)(1,\ldots,1,2)}(z_{1})$. The operator
$R^{(2)}(z_{2})$ is a generic disorder operator, and
$\Phi^{q*}(z_{3})$ with $q^*=0,\pm r^*$ is a singlet or a doublet
$r^*$ operator (we remind that $r^*$ is the $Z_N$ charge in the notation
(\ref{Qnotation})).

The necessary condition for the
above function  to be non-zero is given by the following equation
on the dimensions of the operators $R^{(1)}$, $R^{(2)}$ and $\Phi^{q*}$:
\beq
 \Delta_{R^{(2)}}-\Delta_{\Phi^{q*}}+\frac{1}{(N-1)}
 \left(1-\frac{2}{N}\delta_{q^*,r^*}\right)
 \Delta_{R^{(1)}}=
 (-1)^{\delta_{q^*,r^*}}\frac{N}{N-1}\frac{h^{(2)}_{1}}{h^{(1)}_{1}}
 \Delta_{R^{(1)}}.
 \label{char0}
\eeq
The values  $h^{(1)}_{1}$ and $h^{(2)}_{1}$ are the zero mode
eigenvalues of the operators $R^{(1)}$ and $R^{(2)}$, as defined in
Eq.~(\ref{Rzero1}). While  $h^{(1)}_{1}$  has been  fixed by the
computations presented in Section~2, we do not know the eigenvalue
$h^{(2)}_{1}$ for a generic disorder operator. If we set
$R^{(2)}=\Phi_{(1,\ldots,1,2)(1,\ldots,1)}$ or
$R^{(2)}=\Phi_{(1,\ldots,1)(1,\ldots,1,2)}$, the eigenvalue
$h_{1}^{(2)}$ in Eq.~(\ref{char0}) is known and the dimension
$\Delta_{\Phi^{q*}}$ can be easily calculated for each of the two
channels $q^*=0$ and $q^*=r^*$.

Let us discuss, for example, the case
$R^{(1)}=R^{(2)}=\Phi_{(1,\ldots,1)(1,\ldots,1,2)}$. In this case we
obtain:
\beq
 \Delta_{\Phi^{0}}=\Delta_{(1,\ldots,1)(1,\ldots,1)}=0,
 \qquad \Delta_{\Phi^{r*}}=\Delta_{(1,\ldots,1)(1,\ldots,1,3)}.
\eeq
This  result is in agreement both with the fact that
$\Phi_{(1,\ldots,1)(1,\ldots,1,3)}$ was deduced to be a doublet $r^*$, see
Eq.~(\ref{posconj}), and with the value of the boundary term
$B_{D^{r*}}=(N-1)/8N $. Note also that the
Coulomb gas fusin rules are well respected. Indeed, as discussed previously,
we expect that the fusion $\Phi_{(1,\ldots,1)(1,\ldots,1,2)} \cdot
\Phi_{(1,\ldots,1)(1,\ldots,1,2)}$
produces in the principal channel the operator with
$\vec{\beta}_{(1,\ldots,1)(1,\ldots,1,3)}=
 \vec{\beta}_{(1,\ldots,1)(1,\ldots,1,2)}+
 \vec{\beta}_{(1,\ldots,1)(1,\ldots,1,2)}$.
The $q^*=0$ channel, corresponding to the identity operator, follows the
principal one by a shift
realised by the combination $-\sum_{k=1}^{r} k\vec{e}_{k}$ of
screening vectors.

Actually, even when $h_{1}^{(2)}$ is unknown, we can still do some amount of
verification by using Eq.~(\ref{char0}). Indeed, one of the two channels of
this equation could be used to define $h^{(2)}_{1}$, by assuming a given value
of $\Delta_{\Phi^{0}}$ for instance, chosen at a particular position in the
Kac table. The other channel, with $\Phi^{r*}$, in which enters the same
$h^{(2)}_{1}$, could then serve to check for the presence of
$\Delta_{\Phi^{r*}}$ at the appropriate position in the Kac table, having the
value calculated from the characteristic equation (\ref{char0}). We have in
this way verified the compatibility of the theory with Eq.~(\ref{char0}).

\item $\left \langle \Phi^{r}_{(2)}\Phi^{0}\Phi^{-r}_{(1)}
 \right \rangle$

The module of the operator $\Phi^{-r}_{(1)}$ is supposed to be degenerate at
levels $\delta_{r-1}^{r}$ and $1$, i.e.,
$\Phi^{-r}_{(1)}=\Phi_{(2,1,\ldots,1)(1,\ldots,1)}$ or
$\Phi^{-r}_{(1)}=\Phi_{(1,\ldots,1)(2,1,\ldots,1)}$ (see Section~2).
We derive the following equation for the dimensions:
\beq
 \Delta_{\Phi^{0}}-\Delta_{\Phi^{r}_{(2)}}-\frac{1}{N-1}
 \Delta_{\Phi^{r}_{(1)}}=
 -\frac{N}{N-1}\frac{h^{(2)}_{1}}{h^{(1)}_{1}}\Delta_{\Phi^{r}_{(1)}},
 \label{char1}
\eeq
where $h^{(2)}_{1}$ and $h^{(1)}_{1}$ are respectively the eigenvalues
of the operators $\Phi^{r}_{(2)}$ and $\Phi^{-r}_{(1)}$, as defined in
Eq.~(\ref{zeroeigen}).

We consider the operators $\Phi_{(1,\ldots,1)(2n,1,\ldots,1)}$ and
$\Phi_{(2n,1,\ldots,1)(1,\ldots,1)}$, with $n$ integer.
The above operators have $Z_N$ charge $q=r$, as it can be seen
from Eq.~(\ref{filling}). We know from the reflection-type arguments
that these operators are degenerate at level $\delta_{r-1}^{r}$. If we set
$\Phi^{r}_{(2)}= \Phi_{(1,\ldots,1)(2n,1,\ldots,1)}$  or
$\Phi^{r}_{(2)}= \Phi_{(2n,1,\ldots,1)(1,\ldots,1)}$, then the
correspondent eigenvalue $h^{(2)}_{1}$ is determined by
Eq.~(\ref{fixeig2}). In this case, Eq.~(\ref{char1}) has
respectively the solutions:
\beq
 \Delta_{\Phi^{0}}=\Delta_{(1,\ldots,1)(2n-1,1,\ldots,1)} \mbox{ and }
 \Delta_{\Phi^{0}}=\Delta_{(2n-1,1,\ldots,1)(1,\ldots,1)}.
\eeq
We conclude that the singlets $\Phi_{(2n-1,1,\ldots,1)(1,\ldots,1)}$
and $\Phi_{(1,\ldots,1)(2n-1,1,\ldots,1)}$ are produced in a non principal
channel of the fusion
$\Phi^{r}_{(2)}\cdot\Phi^{-r}_{(1)}$. This is in agreement with the
Coulomb gas rules and with the filling of the weight lattice that we have
given previously.

\item $\left \langle \Phi^{r}_{(2)}\Phi^{1}\Phi^{r}_{(1)}
 \right \rangle.$

In the above function we suppose
$\Phi^{r}_{(1)}=\Phi_{(2,1,\ldots,1)(1,\ldots,1)}$ or
$\Phi^{r}_{(1)}=\Phi_{(1,\ldots,1)(2,1,\ldots,1)}$, and
$\Phi^{r}_{(2)}=\Phi_{(1,\ldots,1)(2n,1,\ldots,1)}$ or
$\Phi^{r}_{(2)}=\Phi_{(2n,1,\ldots,1)(1,\ldots,1)}$.
Taking into account that the modules of these
operators are degenerate at level $\delta_{r-1}^{r}$, we obtain the
characteristic equation:
\beq
 \Delta_{\Phi^{1}}=\Delta_{\Phi^{r}_{(2)}}+\frac{N-3}{N-1}
 \Delta_{\Phi^{r}_{(1)}}.
 \label{char2}
\eeq
Once again, the above equation is consistent with the theory. For
example, if we consider the fusion
$\Phi_{(1,\ldots,1)(2,1,\ldots,1)}\cdot\Phi_{(1,\ldots,1)(2n,1,\ldots,1)}$
we have from Eq.~(\ref{char2}) that
\beq
 \Delta_{\Phi^{1}}=\Delta_{(1,\ldots,1)(2n-1,2,1,\ldots,1)}.
\eeq
The doublet 1 operator $\Phi_{(1,\ldots,1)(2n-1,2,1,\ldots,1)}$
is expected to be produced in a non-principal channel of the
fusion considered above, since
\beq
 \vec{\beta}_{(1,\ldots,1)(2n-1,2,1,\ldots,1)}=
 \vec{\beta}_{(1,\ldots,1)(2,1,\ldots,1)}+
 \vec{\beta}_{(1,\ldots,1)(2n,1,\ldots,1)}-\vec{e}_1.
\eeq
The placing of the doublet $1$ operator at the position
$\vec{\beta}_{(1,\ldots,1)(2n-1,2,1,\ldots,1)}$
is confirmed by the above result.
\end{itemize}

\section{Discussion}

In this paper, we have constructed and analysed the
conformal field theories based on the second solution of the
$Z_N$ parafermionic algebra, with $N=2r+1$ ($r=1,2,\ldots$).

We have achieved the principal goal of giving the Kac formula for these
theories (\ref{Kac}), i.e, the formula for the conformal dimensions of
operators which realise the degenerate representations of the parafermionic
algebra. To obtain this result, we had to determine the boundary terms for
each sector, Eq.~(\ref{Btermsf}), and to determine to which sector belongs each
operator in the Kac table (see Eq.~(\ref{filling})).

We want to stress here that the weight lattice of the Lie algebra $B_r$ is
also known to accommodate the primary fields of the the $WB_{r}$ conformal
field theories. The Kac formula of these theories, was defined, among those of
other $W$ theories, by Fateev and Luk'yanov in Ref.~\cite{luk}. It should
be noticed that the Kac formulae of both the present theory and of $WB_r$
is invariant under the symmetry (\ref{symmetry}).

However, the content of these two conformal theories, $WB_{r}$ and $Z_{2r+1}$,
second solution, are completely different. In the case of the $WB_{r}$ theory,
the $B_r$ weight lattice is filled with two kinds of operators, corresponding
to the Neveu-Schwarz and Ramond sector (alias the $Z_2$ neutral sector and the
spin operator sector). In the case of the $Z_{2r+1}$ theory, second solution,
the same weight lattice accommodates singlet, doublet $q$ and disorder
operators. The important point is that the distribution of singlets, doublet
$q$ and disorder operator on the lattice is consistent with the symmetry
(\ref{symmetry}), as well as with the decoupling of the operators outside the
physical domain (\ref{finite}).

It is natural to assume that, for each $N=2r+1$, the infinite set of unitary
theories with $p=N-2+n$ and $n=1,2,\ldots$ describe multicritical fixed points
in statistical systems with $Z_{2r+1}$ symmetry. It would be interesting to
verify this assumption by finding lattice realisations of the corresponding
models and studying them numerically.

From a theoretical point of view, it is important to remark that it is
not in general (except for some special cases; see Section~2)
practicable to find the dimensions of the fundamental operators by
means of explicit degeneracy computations. We bypassed this problem
by exploiting the properties of the assumed Coulomb gas realisation of
the theory.

This leads us to conjecture that, assuming a particular lattice on which the
primary operators could be accommodated, this kind of approach could greatly
simplify the study of the representations of a given chiral algebra with a
free parameter. The first application of the above assumption would be the
study of the representations of the theory $Z_{N}$, with $N$ even, which
should be based on the $D_{N/2}$ weight lattice \cite{inprep}.

{\bf Acknowledgments:} We would like to thank D.~Bernard, P.~Degiovanni, 
V.~A.~Fateev, and P.~Mathieu for very useful discussions. We are particularly
grateful to M.-A.~Lewis, F.~Merz, and M.~Picco for their collaboration
during the initial stages of this project. \\

\underline{\em Note added in proof.}

After the completion of this work, P.~Furlan attracted our
attention to the paper \cite{Furlan} in which the representation
theory of the chiral algebra (\ref{ope1})--(\ref{ope2}) was considered.
Ref.~\cite{Furlan} also gives partial results for an explicit
free field construction of the coset (\ref{SOcoset}), based on earlier
work of Gepner \cite{Ge}, and of Christe and Ravanini \cite{CR}.
Our work differs from that of Ref.~\cite{Furlan} in classifying
the primary fields representing the coset (\ref{SOcoset}) according to their
$Z_N$ transformation properties, as singlet, $(N-1)/2$ different doublets,
and disorder operators.
The Kac table proposed in Eq.~(4.71) of Ref.~\cite{Furlan}
contains three sectors for each odd $N$, in sharp contrast with
the $(N+3)/2$ sectors identified in the present work. Accordingly,
we do not agree on the dimensions of primary operators proposed
in Ref.~\cite{Furlan}.


\begin{thebibliography}{99}

\bibitem{para1}   V.~A.~Fateev and A.~B.~Zamolodchikov,
                  Sov.~Phys.~JETP {\bf 62}, 215 (1985).
\bibitem{para2_3} V.~A.~Fateev and A.~B.~Zamolodchikov,
                  Theor.~Math.~Phys.~{\bf 71}, 451 (1987).
\bibitem{para2}   Vl.~S.~Dotsenko, J.~L.~Jacobsen and
                  R.~Santachiara,
                  Nucl.~Phys.~B {\bf 656}, 259 (2003).
\bibitem{goddard} P.~Goddard and A.~Schwimmer,
                  Phys.~Lett.~B {\bf 206}, 62 (1988).
\bibitem{dis1}    V.~A.~Fateev and A.~B.~Zamolodchikov,
                  Sov.~Phys.~JETP {\bf 63}, 913 (1986).
\bibitem{Felder}  G.~Felder, Nucl.~Phys.~B {\bf 317}, 215 (1989);
                  Nucl.~Phys.~B {\bf 324}, 548 (1989).
\bibitem{Walg}    P.~Bouwknegt, J.~McCarthy and K.~Pilch,
                  Lect.~Notes Phys.~{\bf M42}, 1--204 (1996).
\bibitem{luk}     V.~A.~Fateev and S.~I.~Luk'yanov,
                  Sov.~Sci.~Rev.~A.~Phys.~{\bf 15}, 1--117 (1990).
\bibitem{inprep}  Vl.~S.~Dotsenko, J.~L.~Jacobsen and
                  R.~Santachiara, in preparation.
\bibitem{Furlan}  P.~Furlan, R.R.~Paunov and I.T.~Todorov,
                  Fortschr.~Phys.~{\bf 40}, 211 (1992).
\bibitem{Ge}      D.~Gepner, Nucl.~Phys.~B {\bf 290}, 10 (1987).
\bibitem{CR}      P.~Christe and F.~Ravanini,
                  Int.~J.~Mod.~Phys.~A {\bf 4}, 897 (1989).
\end{thebibliography}
\end{document}